\documentclass[10pt]{elsarticle}

\usepackage{amssymb}
\usepackage{url}
\usepackage{amsthm}

\usepackage{graphicx}
\usepackage{caption}
\usepackage{subcaption}
\usepackage{color}
\usepackage{mathtools}
\usepackage{amsfonts}
\newcommand{\vect}[1]{\boldsymbol{\mathbf{#1}}} 
\journal{}

\begin{document}

\begin{frontmatter}



\title{Isogeometric finite element analysis of time-harmonic exterior acoustic scattering problems}

 \author[label1]{Tahsin Khajah}
 \author[label2]{Xavier Antoine}
 \author[label3,label4,label5]{St\'{e}phane P.A. Bordas}
 \address[label1]{Department of Mechanical Engineering, University of Texas at Tyler, USA.}
 \address[label2]{Institut Elie Cartan de Lorraine, Universit\'e de Lorraine, Inria Nancy-Grand Est, SPHINX team, F-54506
  Vandoeuvre-l\`es-Nancy Cedex, France.}
 \address[label3]{Universit\'{e} du Luxembourg, Legato-Team, Institute of Computational Engineering}
 \address[label4]{Cardiff University, Institute of Mechanics and Advanced Materials }
 \address[label5]{University of Western Australia, Intelligent Systems for Biomedicine Laboratory}
 


\address{}

\begin{abstract}
We present an isogeometric analysis of time-harmonic exterior acoustic problems. The infinite space is truncated by a fictitious boundary and (simple) absorbing boundary conditions are applied. The truncation error is included in the exact solution so that the reported error is an indicator of the performance of the isogeometric analysis, in particular of the related pollution error. Numerical results performed with high-order basis functions (third or fourth orders) showed no visible pollution error even for very high frequencies. This property combined with exact geometrical representation makes isogeometric analysis a very promising platform to solve high-frequency acoustic problems. 
\end{abstract}


\end{frontmatter}

\tableofcontents


\section{Introduction}
\label{Intro}
The purpose of this paper is to investigate the performance of Non-Uniform Rational B-Splines (NURBS)-based 
 {\it Isogeometric Analysis (IGA)} finite element methods (IGAFEM) for solving  time-harmonic acoustic wave scattering and acoustic wave propagation problems, most particularly for small wavelengths.
To fix the notations, let us consider $\Omega^- $ as a $d$-dimensional ($d=1, 2, 3$) scattering bounded domain of $\mathbb{R}^{d}$,
with shape $\Gamma:=\partial\Omega^-$. We introduce the associated exterior (i.e. unbounded) domain
of propagation $\Omega^+:=\mathbb{R}^{d}/\overline{\Omega^-} $. 
Then, solving the scattering problem leads to computing the wave field $u$, solution to the Boundary-Value Problem (BVP): 
given an incident (plane wave) field $u^{\textrm{inc}}$, find $u$ such that
\begin{equation}\label{systHelm}
\begin{aligned}
&\Delta u + k^2 u = 0, \quad \text{in} \quad \Omega^+, \\
&\partial_{\vect{n}_\Gamma} u = \vect{g} := -\partial_{\vect{n}_\Gamma} u^{\textrm{inc}} 
, \quad \text{on} \quad \Gamma, \\
&\lim_{|\vect{x}|\to +\infty} |\vect{x}|^{(d-1)} \Big( \nabla u \cdot \frac{\vect{x}}{|\vect{x}|}-iku \Big) = 0,
\end{aligned}
\end{equation}
where $\Delta$ is the Laplacian operator, $\nabla$ the gradient operator 
and $ \vect{n}_{\Gamma}$ is the outward-directed unit normal vector to $\Omega^-$. The spatial
variable is $\vect{x}=(x,y)$ in 2D and $\vect{x}=(x,y,z)$ in 3D.
The wavenumber $k$ is related to the wavelength $\lambda$ by the relation: $\lambda:=2\pi/k$.
The boundary condition is a  Neumann (sound-hard) boundary condition
(but other boundary conditions could also be considered).
Denoting by $ \vect{a} \cdot \vect{b}$  the hermitian inner-product of two complex-valued vector fields $\vect{a}$ and $\vect{b}$,
then the last equation of system (\ref{systHelm})  is known as the Sommerfeld's radiation condition \cite{colton1983integral, NedelecBook} at infinity, which presents the outgoing wave to the domain.
Various numerical methods have been developed to accurately solve the exterior scattering problem,
e.g.  finite elements \cite{Ihlenburg1998,IhlenburgBabuskaI97,BouillardIhlenburg99,Beriot16,GabardGamallo11,Ortiz2001,ThompsonPinsky95,GiorgianiModesto13,Thompson2006},
boundary elements \cite{colton1983integral,NedelecBook,Bruno2004,ChandlerWildeReview,FMMBook,AntoineDarbasQJMAM,AntoineDarbasESAIM,DDL}, 
enriched and various modified (wave-based, hybrid, asymptotic)
finite elements \cite{FarhatEnriched,CessenatDespresUltraWeak,GabardGamallo,HarariMagoules,Geuzaine2008,AntoineCG2009,Huttunen2008,LAGHROUCHE2000,Kechroud2009,GILADI2001,ASTLEY1983}.
A challenging and still outstanding question for numerics and applications is related to the so-called \textit{high-frequency} regime,
where the wavelength $\lambda$ is very small compared with the characteristic length of the scatterer $\Omega^{-}$ 
\cite{Thompson2006,Babuska1997}.

A first ``exact'' method consists in writing an integral equation on the surface $\Gamma$ to represent the exterior field, and then solving this
by means of a boundary element method, combined with fast evaluation algorithms (e.g. Fast Multilevel Multipole Methods
(MFMM)�\cite{FMMBook,DDL,Darve}, Adaptive Cross Approximation (ACA) \cite{Bebendorf,ZhaoVouvakis}) and preconditioned (matrix-free) 
Krylov subspace iterative solvers
(GMRES) \cite{AntoineDarbasQJMAM,AntoineDarbasESAIM,SaadBook}. This approach has the very interesting property to reduce  the initial problem to a finite
$(d-1)$-dimensional problem (set on $\Gamma$) and to be relatively stable thanks to the frequency regime (typically between
$n_{\lambda}=5$ to $10$ points per wavelength are used, with $n_{\lambda}=\lambda/h$, and $h$ the meshsize).
 Nevertheless, the method is nontrivial to adapt to complex geometries,
when the boundary conditions are modified or when a high-order of accuracy is required for the solution. In particular, 
developing  efficient algorithms when the geometry is described
with high accuracy remains an open question, most particularly for large wave numbers $k$ since
the number of degrees of freedom is potentially very large. 

Another standard and very attractive engineering-type 
approach consists in truncating the exterior domain $\Omega^{+}$ through the introduction of a fictitious
boundary/layer $\Sigma$ that surrounds the scatterer $\Omega^{-}$, resulting in a bounded computational domain $\Omega_{b}$
with inner boundary $\Gamma$ and outer boundary/layer $\Sigma$.
When $\Sigma$ is a boundary, an {\it Absorbing Boundary Condition (ABC)}  \cite{ABCReview1,ABCReview2,ABCReview3,
AntoineBarucqBendali,AntoineABC,Turkel1,Turkel2,Turkel3} is set on the fictitious boundary
 which introduces a truncation error even at the continuous level.
 This can be reduced by considering high-order (local) ABCs. A very popular alternative is, rather, to consider
  a surrounding absorbing layer and modifying the nonphysical media to obtain a Perfectly Matched Layer
   \cite{BerengerPML,ChewPML,BermudezPML,BermudezReview}.

In this paper, since our goal is to understand the quality of the NURBS-approximations used in Isogeometric Analysis (IGA) to approximate the wave field,
we will consider simple boundary conditions where we can separate the continuous truncation error from the numerical approximation. 
 High-order ABCs as well as PML will be treated in future work.
When the problem is truncated, any numerical method can be used such as e.g.  finite-element methods, the resulting
linear system being then solved through well-adapted domain decomposition techniques
\cite{despres-etal:92,despres:90,GanderMagouNataf,BoubendirAntoineGeuzaine}.
In the high frequency regime, it is well-known that
the numerical solution suffers from a phase shift due to numerical dispersion which is called the {\it pollution error}
 \cite{Babuska1997,Ihlenburg1995,Ihlenburg1997,Bouillard1999} of the finite element method. 
 To maintain a prescribed pollution error, the mesh density $n_{\lambda}$
 should be increased faster than the wave number leading to high computational cost for high frequency problems. Therefore, the FEM is limited to a upper frequency bound for which the computational cost becomes prohibitive and
  increasing the order of the polynomial basis functions used in conventional FEM is required to
 reduce the pollution error, but it does not fully eliminate it.

 Original  techniques were recently proposed to reduce the pollution error.
 One approach is to replace the basis functions, usually a polynomial, with plane wave
  functions \cite{LAGHROUCHE2000,GILADI2001}. Since pollution error can be viewed as a phase shift, another
  hybrid approach is to asymptotically approximate the phase of the  solution
   and to reformulate the problem into an equivalent problem with a slowly varying unknown \cite{Bruno2004,Geuzaine2008,AntoineCG2009,ASTLEY1983,Astley1983-1,baumeister1974,baumeister1977,Turkel2004}. 
Nevertheless, until now, the only viable solution for complex problems
is to consider a sufficiently high-order polynomial basis into the FEM together with high-order meshes to represent accurately
 the geometry $\Gamma$. 
 
The aim of the present paper is to analyze 
the quality of alternative approximation methods based on B-splines and NURBS-based  {\it Isogeometic Analysis (IGA)} and to compare the properties of these approximation schemes to more traditional, Lagrange-based high-order FEM. The goal of the paper is, specifically, to investigate the properties of the methods for large wave numbers $k$. Isogeometric analysis (IGA) was introduced over 10 years ago to streamline the transition from Computer-Aided Design to Analysis  \cite{Hughes2005,Cottrell2009}. The central idea of the approach is to use the same shape functions to approximate the field variables as those used to describe the geometry of the domain. As the geometry of computational domains is typically provided by Computer-Aided Design (CAD) software, it is sensible to use NURBS shape functions, which are the most commonly used functions in CAD. 
 
Since 2005, IGA has been developed within the finite element, 
boundary element \cite{Simpson2012,Lian2013,Simpson2013} and was enriched through partition of unity \cite{Nguyen2015} 
IGA benefits from exact (and smooth) geometry representation and no loss of boundary details. This is fundamental in problems involving wave propagation or turbulence, as spurious wave scattering and vortices typically appear close to rough boundaries. 
IGA therefore alleviates the meshing burden as if the CAD model is modified, any modifications are immediately inherited by the shape functions used for analysis. When coupled to boundary element approaches, this suppresses completely the generation and regeneration of the mesh without any regeneration of the mesh \cite{Lian2013,Lian2016}. When used within a finite element approach, volume parameterization is still required, but the shape optimization process is simplified \cite{Wolfgang2008, Fußeder2015}.
 Since CAD models are generated from the boundary data, {\it IGA Boundary Element Method (IGABEM)} is  promising in bridging the gap between CAD and analysis offering a unique shape optimization package \cite{Scott2013, Coox2016, Kostas2015, Lian2013R}. 

IGA additionally offers higher inter-patch continuity, large support sizes and superior refinement capabilities compared to Lagrange-based approximation schemes. A recent review of IGA and its computer implementation aspects is available in \cite{Nguyen2015}. GEOPDE is another open source IGA research tool \cite{deFalco2011} which provides a flexible coding structure. The application of IGA and its performance in solving one- and two-dimensi\-onal eigenvalue and interior Helmholtz problems have been studied \cite{Hughes2014290,Hughes20084104, Cottrell20065257, Coox2016441} in which IGA showed superiority to conventional FEM. However, the performance of IGA in solving exterior scattering problems and related pollution error is still not thoroughly investigated.  

In this paper, we study the  performance of IGA for  solving time-harmonic exterior scattering problems in one, two and three dimensions. First, in Section \ref{sec:IGA}, we briefly present the Isogeometric Analysis and the related NURBS and B-Spline basis functions. 
Then in Section \ref{sec:1D}, we consider a simple but meaningful one-dimensional scattering problem and provide a pseudo-code to solve it in IGA context. 
In Section \ref{sec:2D}, we consider two-dimensional problems and study the performance of IGA in estimating the scattered field of a duct and a circular cylinder. 
The in-house IGA and Lagrange-based FEM codes developed in this study were written in Matlab {\textregistered} following the GEOPDE platform for two- and three-dimensional examples. 
Finally, section \ref{SectionConclude} conclude the paper.
%
%
%
%
\section{Isogeometric Analysis (IGA)}
\label{sec:IGA}
In IGA, both the physical model and the solution space are constructed by {\it B-spline/NURBS} functions. The model is constructed using a set of control points and {\it knot vectors} in each spatial direction. A knot vector is defined as a set of non-decreasing coordinates and represented by $ \vect{\xi}=\{\xi_1,\xi_2,...,\xi_{(n+p+1)}\} $, where $ \xi_i $ is the $ i^{th} $ knot and $ i$ is the knot index, $ i=1,2,\cdots,n+p+1$,
 where $p$ is the polynomial order and $ n $ is the number of basis functions. Usually an {\it open knot vector} is used in IGA where the first and the last knot values appear $ p+1 $ times. The elements in IGA are defined as non-zero knot intervals.  If the knots are spaced equally within the knot vector, the knot vector is called {\it uniform} otherwise it is {\it non-uniform}. 
\subsection{B-Splines}
\begin{figure}[!hbt]
    \centering
        \begin{subfigure}[b]{0.49\textwidth}
        \includegraphics[width=\textwidth]{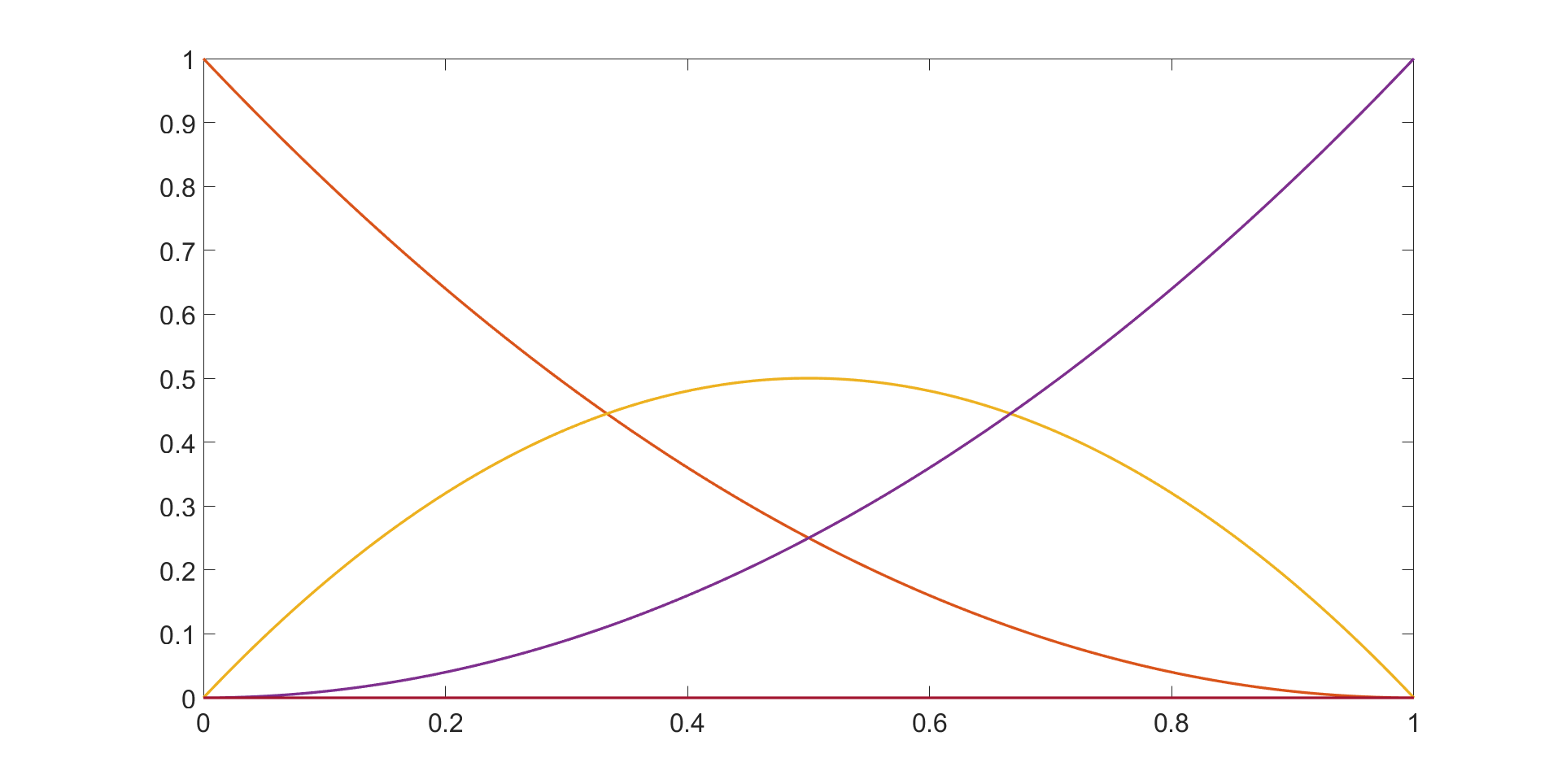}
        \caption{$p=2$}
        \label{fig:p2}
    \end{subfigure} 
    \begin{subfigure}[b]{0.49\textwidth}
        \includegraphics[width=\textwidth]{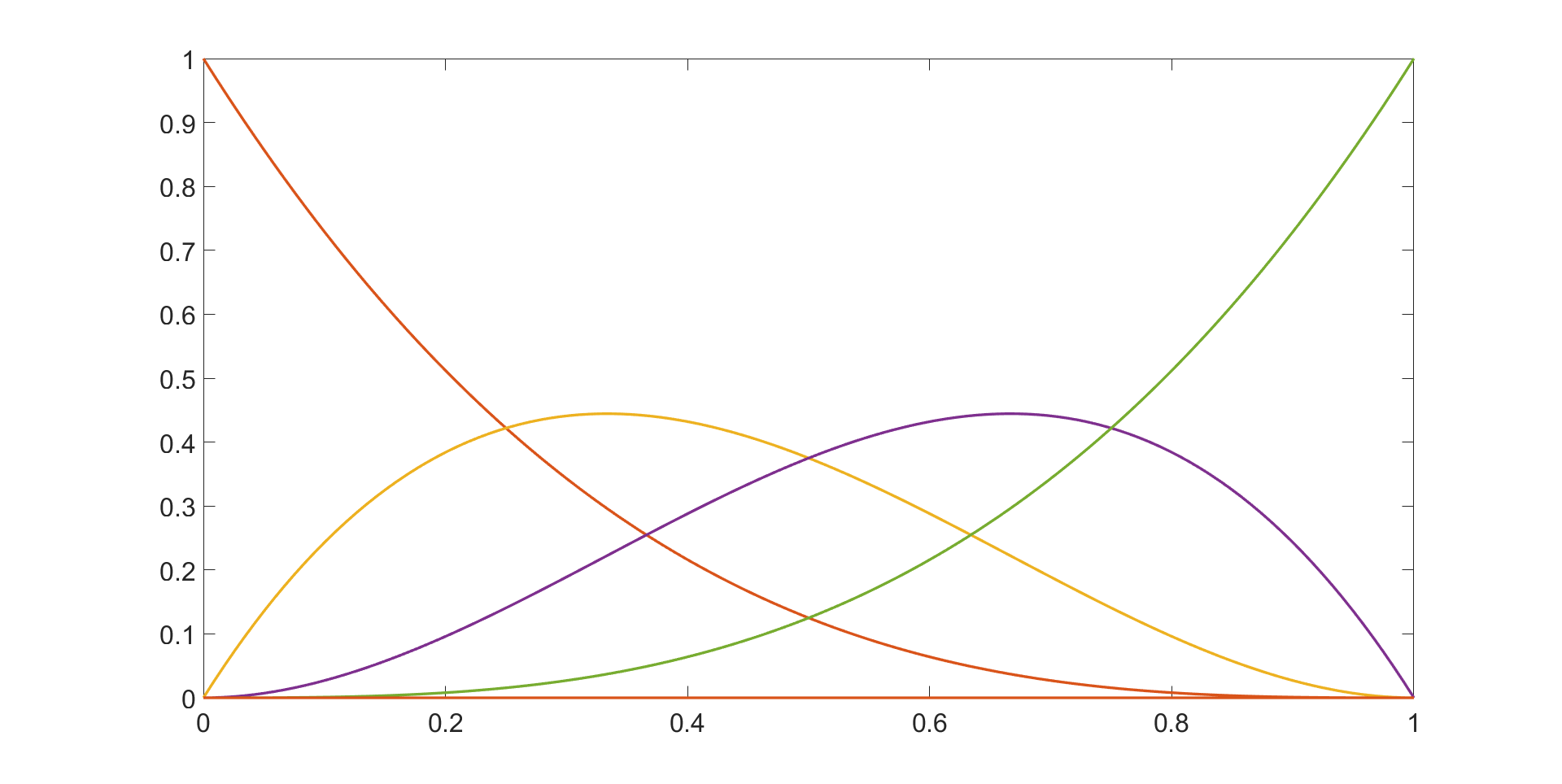}
        \caption{$p=3$}
        \label{fig:p3}
    \end{subfigure}\\
      \begin{subfigure}[b]{0.49\textwidth}
        \includegraphics[width=\textwidth]{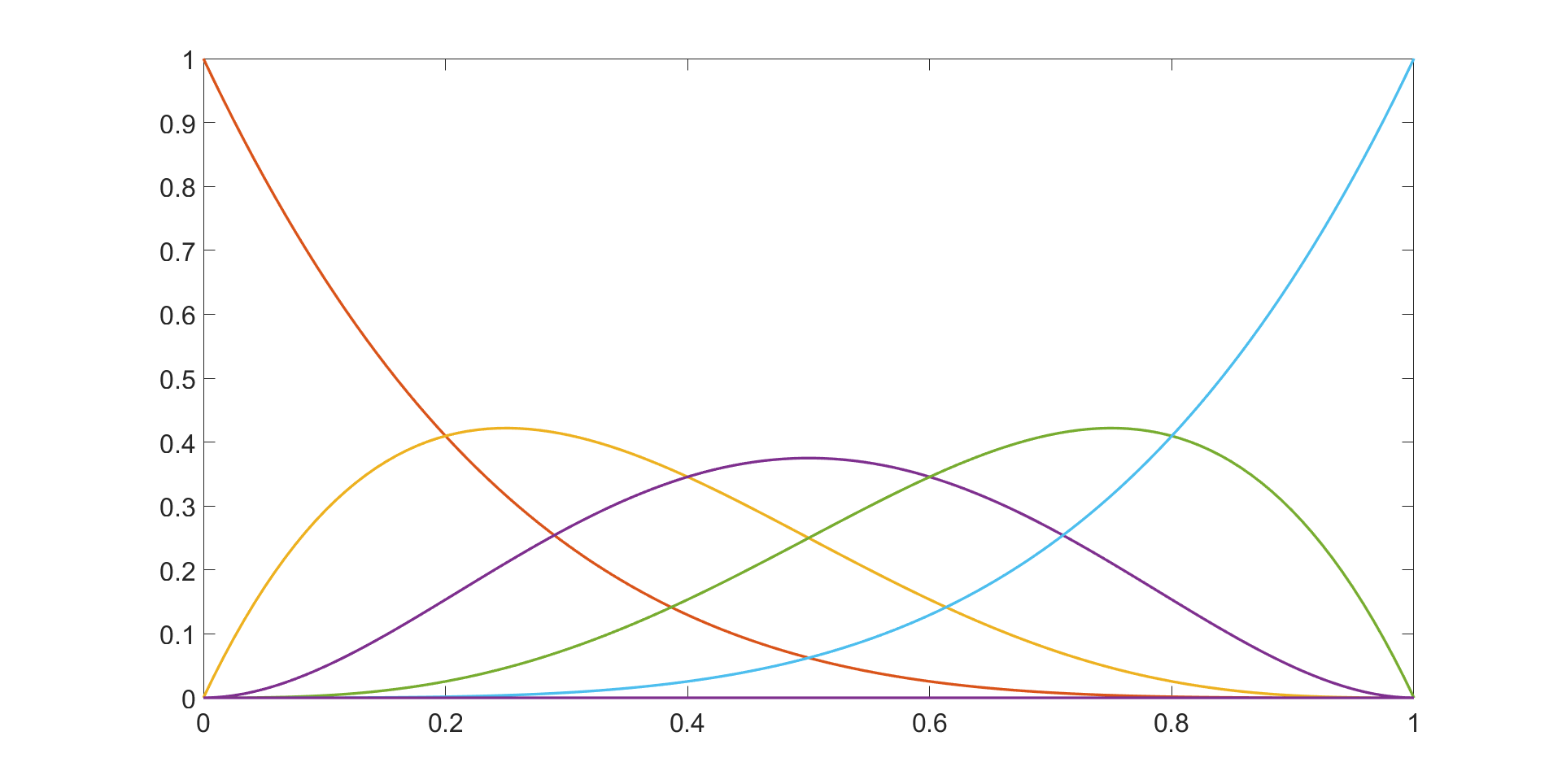}
        \caption{$p=4$}
        \label{fig:p4}
    \end{subfigure} 
    \begin{subfigure}[b]{0.49\textwidth}
        \includegraphics[width=\textwidth]{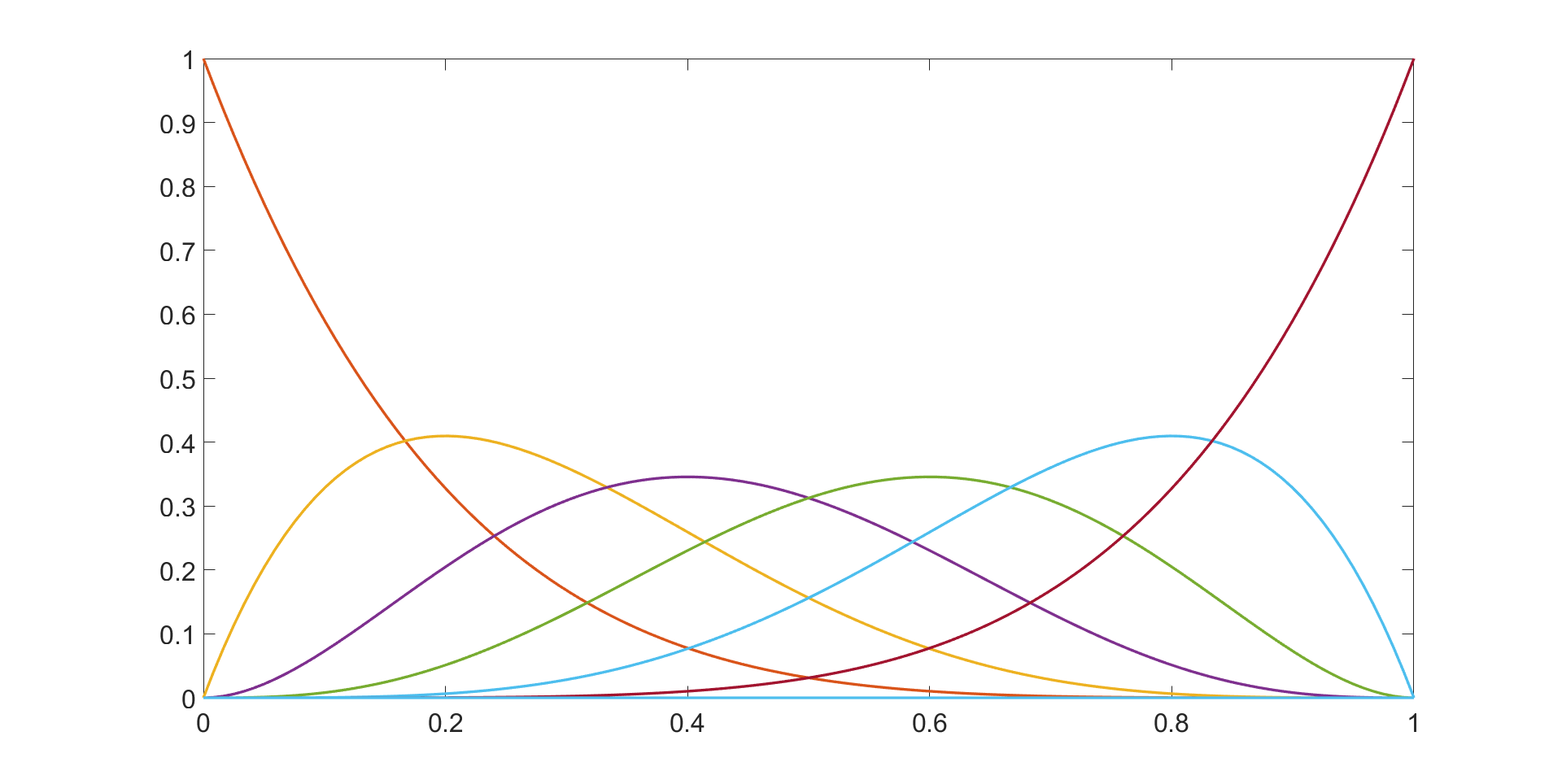}
        \caption{$p=5$}
        \label{fig:p5}
 \end{subfigure}
  \caption{The B-Spline basis functions, $p=1 \cdots, p=5$.}
\label{fig:BSpline_basis}
\end{figure}

\begin{figure}[!hbt]
\centering
\includegraphics[width=\textwidth]{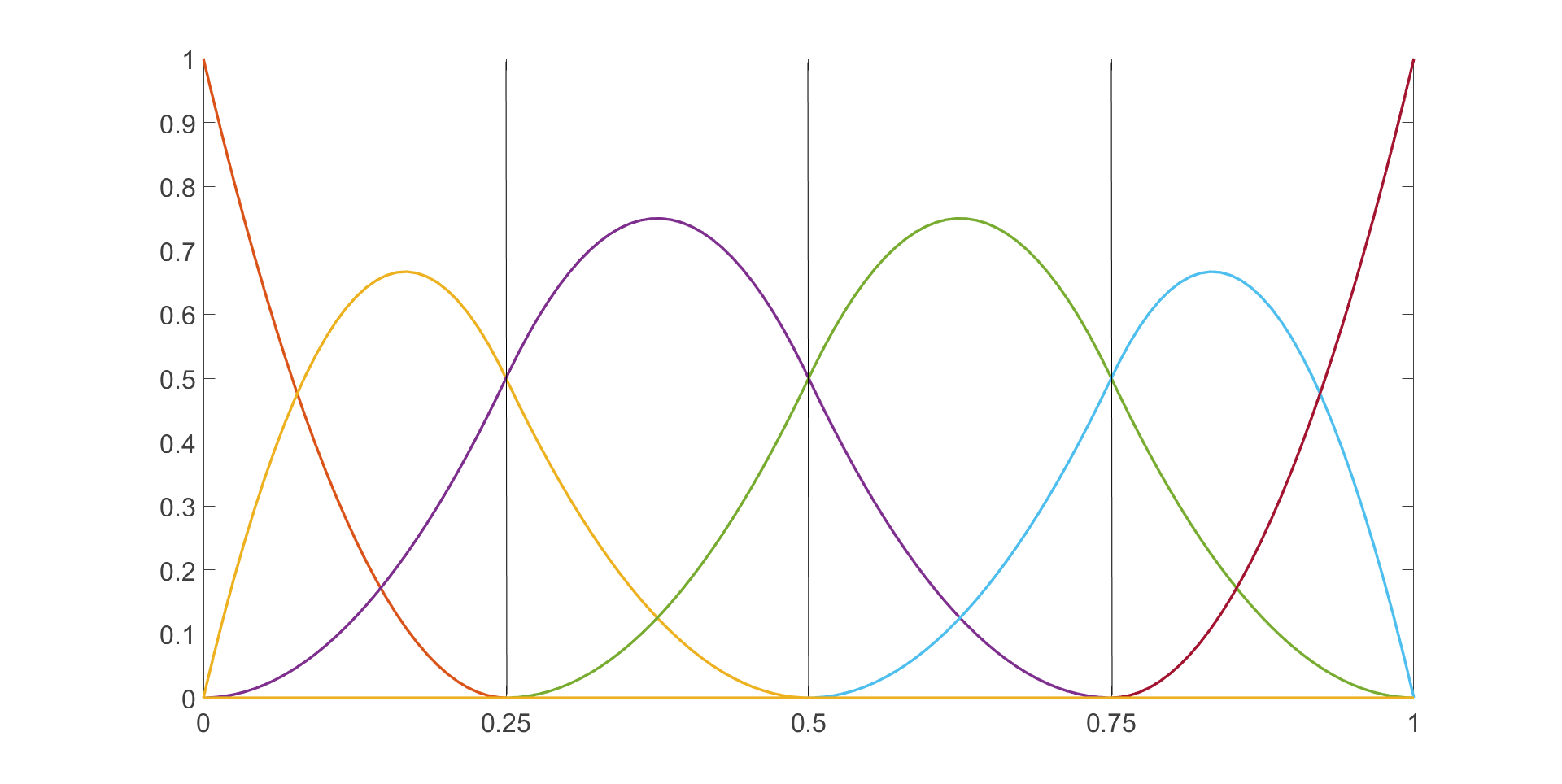}
\caption{The quadratic ($p=2$) basis function of a 1D domain defined over $ \vect{\xi} = \{0, 0, 0, 1/4, 1/2, 3/4, 1, 1, 1\}$ and representing four elements.}
\label{fig:4ep2basis}
\end{figure}

%
A B-Spline basis function is defined recursively by the Cox-de Boor recursion formula starting with the zeroth order $ (p=0) $ basis function:
\begin{equation}
\hspace*{-6.5cm}  \text{for} \ p = 0, \ N_{i,p}(\xi) = 
\left \{ \begin{aligned}
         1 & \ \ \ \xi_i \leq \xi \leq \xi_{i+1},\\
         0 & \ \ \ \text{otherwise},                                                                                                                                                                                                                            \label{eq.Bsp1}
         \end{aligned}
\right.
\end{equation}
\begin{equation}
\hspace*{-.254cm} \text{for}\ p = 1, 2, 3, \cdots \ \ \ N_{i,p}(\xi) = \frac{ \xi - \xi_i}{\xi_{i+p}-\xi_i}N_{i,p-1}(\xi) + \frac{ \xi_{i+p+1} - \xi}{\xi_{i+p+1}-\xi_{i+1}}N_{i+1,p-1}(\xi),          \label{eq.Bsp2}
\end{equation}
where $0/0 $  is defined to be zero. To compute the tangent and the normal functions, it is required to calculate the derivative of B-Splines. Efficient algorithms to calculate B-Splines and their derivatives can be found in the NURBS Book by Piegl and Tiller \cite{Piegl97}. The first-order B-Spline functions are identical to their Lagrangian (FEM) counterparts. Also, the B-Spline�s number of required shape functions for a specific order is similar to that of the FEM. For example, three shape functions are required to construct second-order Lagrangian or B-Spline shape functions. In contrast to FEM, the shape functions in IGA are non-negative. The first derivative of the B-spline basis function is calculated using the following recursive formula:
\begin{equation}
\frac{d}{d \xi} N_{i,p} (\xi) = \frac{p}{\xi_{i+p} - \xi_i} N_{i,p-1}(\xi)- \frac{p}{\xi_{i+p+1} - \xi_{i+1}} N_{i+1,p-1}(\xi).
\end{equation}
These functions exhibit $C^{p-m_i } $ continuity across knot $\xi_i $  where $ m_i $ is the multiplicity of the knot $\xi_i $. 
 The B-Spline basis function corresponding to $p=1, \cdots, p=5$ are shown in Fig. \ref{fig:BSpline_basis} where the knot vectors are defined to represent one element (knot span)
\begin{equation}
  \boldsymbol{\xi} =\{ 
    \underbrace{0, \cdots, 0}_\text{p+1 times}, \cdots, \underbrace{1, \cdots, 1}_{\text{p+1 times}} 
\}.
\end{equation}
The quadratic B-Spline basis function for a 1D domain divided into four elements is shown in Fig. \ref{fig:4ep2basis} where the knot vector is defined as $$ \vect{\xi} = \{0, 0, 0, 1/4, 1/2, 3/4, 1, 1, 1\}.$$ The four non-zero knot spans correspond to the four elements of this one-dimensional domain.

\subsubsection{B-Spline curves }
The piecewise-polynomial B-Spline curves in $ \mathbb{R}^d $  are constructed as a linear combination of B-Spline basis functions:
\begin{equation}
\mathbf {C}(\xi) = \sum_{i=1}^{n} \vect{N}_{i,p} (\xi) \vect{B}_i.                                                                                                                                                                                \label{eq:Bcurve}
\end{equation}
The vector-valued coefficients of the basis functions in Eq.\ref{eq:Bcurve} are control points  ($ \mathbf{B}_i \in \mathbb{R}^d , i=1,2\cdots , n $). Since $ \mathbf{C}(\mathbf{\xi}) $ is representing a curve, the corresponding control points, $ \mathbf{B}_i $, are analogous to nodal coordinates in FEM. The resulting curve is $ C^{p-1} $-continuous everywhere except at the location of the repeated knots with $ m_i $  multiplicity, where it is $ C^{p-m_i } $-continuous. Therefore, when utilizing open knot vectors, the resulting curve is $ C^0 $-continuous, tangent to the control polygon, and interpolatory at the start and the end of the curve and where the multiplicity is equal to the polynomial order. 
The support of a B-Spline basis function is local. In other words, moving a single control point can affect the geometry of no more than $ p+1 $ elements of the curve.
\subsubsection{B-Spline surfaces and volumes}
A tensor product B-Spline surface is defined as:
\begin{equation}
\vect{S}(\xi , \eta) = \sum_{i=1}^{n}\sum_{j=1}^{m}N_{i,p}(\xi)M_{j,q}(\eta) \vect{ B}_{i,j},                                                                                                                                              \label{eq.98}
\end{equation}
where $ {\vect B}_{i,j}, i=1,2,\cdots,n, \  j=1,2,\cdots, m $  are control points, and $ N_{i,p} (\xi) $  and $ M_{j,q} (\eta) $ are univariate B-Spline functions of orders $ p $ and $ q $ corresponding to the knot vectors $ \vect{\Xi}^1=\{\xi_1,\xi_2,\cdots, \xi_{n+p+1} \} $ and  $ \vect{\Xi}^2 = \{\eta_1,\eta_2,\cdots,\eta_{m+q+1}\} $, respectively.
Similarly, a B-Spline volume is constructed as a tensor product of three B-Spline basis functions:
\begin{equation}
\vect{V}(\xi , \eta, \zeta) = \sum_{i=1}^{n}\sum_{j=1}^{m}\sum_{k=1}^{l}N_{i,p}(\xi)M_{j,q}(\eta)L_{k,r}(\zeta) \vect{B}_{i,j,k}.                                                                                 \label{eq.99}
\end{equation}
%
\subsection{NURBS}
There are certain geometries, such as circles and ellipsoids, which B-Spline functions and Lagrange polynomials cannot represent exactly. In the CAD community, B-Spline functions are replaced with a more general form to overcome this shortcoming. This new form is called {\it Non-Uniform Rational B-Splines (NURBS)}. NURBS shape functions are defined as:
\begin{equation}
R_i^p(\xi) = \frac{N_{i,p}(\xi)w_i}{W(\xi)} = \frac{N_{i,p}(\xi)w_i}{\sum_{i=1}^{n} N_{i,p}(\xi) w_i},                                                                                                                                    \label{eq:NURBS}
\end{equation}
where $ \{N_{i,p} \}$ is a set of B-Spline basis functions and $ \{w_i \} $ is a set of positive NURBS weights. By appropriate weight selection both polynomials and circular arcs can be described. NURBS curves are defined as follows:
\begin{equation}
\vect{C}(\xi) = \sum_{i=1}^{n} R_i^p(\xi) \vect{B}_i.                                                                                                                                                                                                                              \label{eq:NURBScurve}
\end{equation}
Similarly, rational surfaces and volumes are defined in terms of rational basis functions as:
\begin{equation}
R_{i,j}^{p,q} (\xi, \eta) = \frac{ N_{i,p}(\xi) M_{j,q}(\eta) w_{i,j}}{\sum_{i=1}^{n} \sum_{j=1}^{m} N_{i,p} (\xi) M_{j,q}(\eta) w_{i,j}},                                                                             \label{eq:NURBSsurf}
\end{equation}
\begin{equation}
R_{i,j}^{p,q} (\xi, \eta, \zeta) = \frac{ N_{i,p}(\xi) M_{j,q}(\eta) L_{k,r}(\zeta) w_{i,j,k}}{\sum_{i=1}^{n} \sum_{j=1}^{m}\sum_{k=1}^{l} N_{i,p} (\xi) M_{j,q}(\eta) L_{k,r}(\zeta) w_{i,j,k}}.                                                                      \label{eq:NURBSvol}
\end{equation}
If the weights are all equal, NURBS basis functions will reduce to their B-Spline counterparts and the corresponding curve becomes a non-rational polynomial again. The geometry of an element in IGA can be expressed as:
\begin{equation}
\vect{x}^e(\tilde{ \xi}) = \sum_{a=1}^{n_{en}} \vect{P}_a^e R_a^e (\tilde{\xi}),
\end{equation}
where $n_{en} = (p+1)^{d_p}$ and $d_p$ is the spatial dimension. The field $\vect {u}(\vect{x})$ is similarly expressed as:
\begin{equation}
\vect{u}(\tilde{\xi}) = \sum_{a=1}^{n_{en}} \vect{d}_a^e R_a^e (\tilde{\xi}),
\end{equation}
where $ \boldsymbol{d}_a^e R_a^e$ is the set of control variables. As a consequence of the parametric definition of basis functions in IGA, it is necessary to consider two mapping for integration
\begin{equation}
\begin{array}{l}
\displaystyle \int_\Omega f(\vect{x}) d \Omega = \sum_{e=1}^{n_{el}}\int_{\Omega^e}f(\vect{x})d\Omega  
\displaystyle  =\sum_{e=1}^{n_{el}}\int_{\hat{\Omega}}^{e} f (\vect{x} (\vect{\xi})) | J_{\vect{\xi}}| d \hat{\Omega}  \\�\quad\quad\quad
\quad\quad
\displaystyle  =\sum_{e=1}^{n_{el}}\int_{\tilde{\Omega}} f (\vect{x}(\tilde{\phi}^e (\vect{\tilde{\xi}})) | J_{\vect{\xi}}| | J_{\tilde{\vect{\xi}}} |  d \tilde{\Omega},  \\
\end{array}
\end{equation}
where $\tilde{\Omega} $ and $\hat{\Omega} $ are representing the parent and parameter spaces respectively, and $\Omega $ is the physical space. 
\section{A one-dimensional toy scattering problem}
\label{sec:1D}
Now, let us come back to our physical problem by  considering the simple one-dimensional Neumann scattering
 problem solved in \cite{AntoineCG2009}:
\begin{equation}
\begin{aligned}
& \partial_x^2 u +k^2u = 0, \quad \ \ &\text{in} \ \ \Omega_b = (0,1), \\
& \partial_{x} u = ik, \quad &\text{at} \quad \Gamma = \{0\}, \\
& \partial_x u -i ku = 0, \quad &\text{at} \quad \Sigma = \{1\}.
\label{eq:1dBVP}
\end{aligned}
\end{equation}
The variational formulation is given as:
\begin{equation}
\int_{\Omega_b} \{\partial_x u \partial_x \bar{v} - k^2 u \bar{v} \} dx -ik(u\bar{v})(1)=-(g\bar{v})(0), \quad \forall v  \in  H^1(\Omega_b).
\end{equation}
The solution of the one-dimensional BVP  (\ref{eq:1dBVP}) is: $u^{\textrm{ex}}(x) = e^{ikx}$, which represents the scattering of an incident plane wave by the left half space. An exact transparent boundary condition is considered for the fictitious boundary at $\Sigma = \{1 \}$ using the Dirichlet-to-Neumann operator $\Lambda = ik$ on $\Sigma$, resulting in the transparent
boundary condition $ \partial_x u = \Lambda u$. 
%
%

\begin{figure}[!hbt]
\centering
\includegraphics[width=1\textwidth]{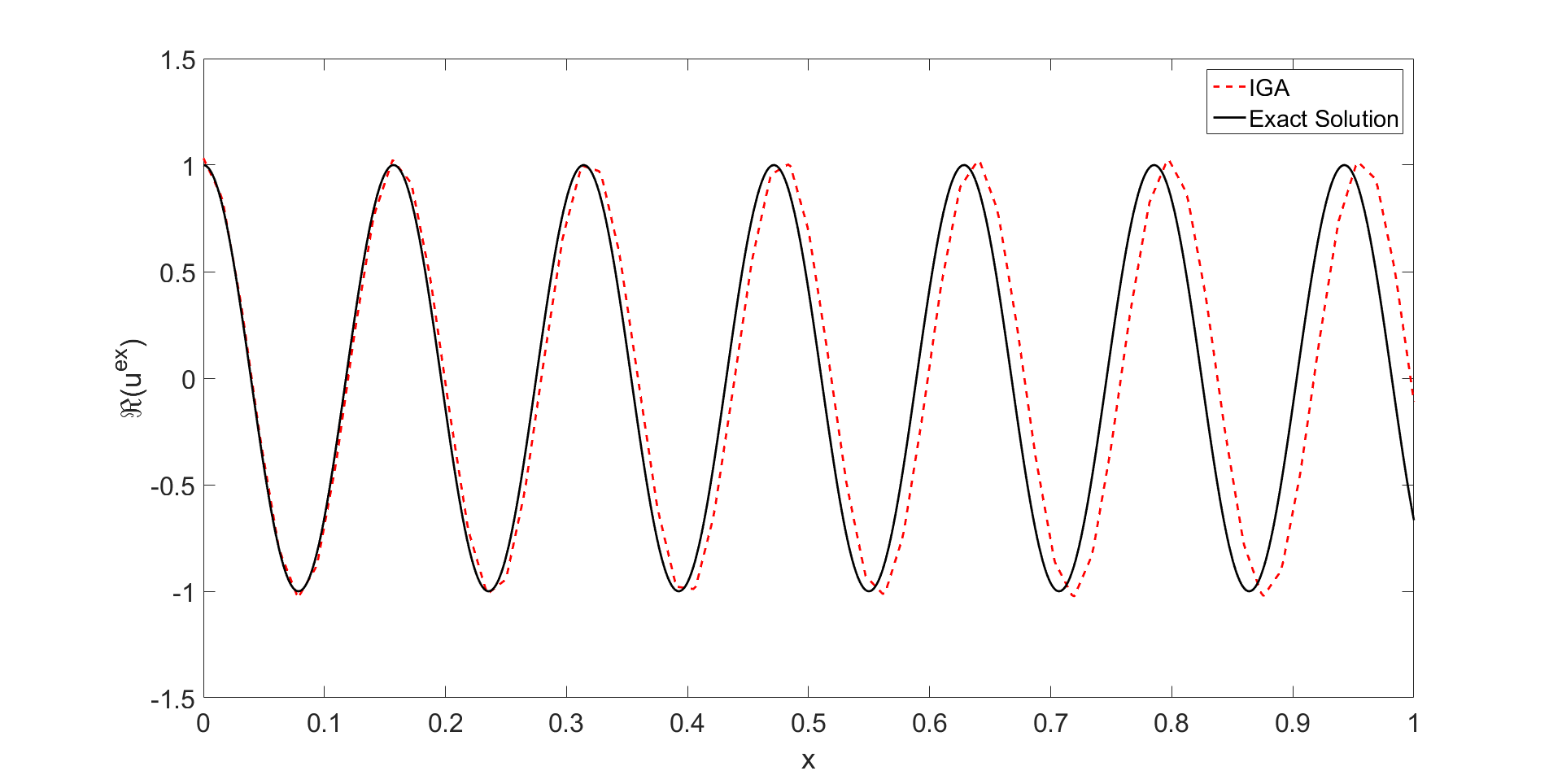}
\caption{The pollution error is the numerical solutions $\Re(u_{h})$ of IGA ($p=1$), is visible when compared with the exact solution $\Re(u^{\textrm{ex}})$, for $k=40$ and $n_{\lambda}=10$.}
\label{fig:IGAFEM}
\end{figure}

It is shown that the Finite Element approximated solution of the BVP described in system (\ref{eq:1dBVP}) suffers from the pollution error \cite{Ihlenburg1997,AntoineCG2009,Thompson2006,Ihlenburg1998}.
The pollution error has a direct relation with the wavenumber $k$ and  increases with frequency. It is the polynomial basis used in conventional FEM which is inadequate to represent the wavefield. Since the main difference of the IGA with the conventional FEM is in the selection of the basis function, it is interesting to study the performance of IGA specially in high frequencies and its pollution error. We divide the computational bounded domain $\Omega_b$ into $N_h$ uniform non-zero knot spans (elements) so that each segment length is $1/h$. We define the density of discretization as the number of elements per wavelength and denote it with $n_{\lambda} = \lambda/h$. A Matlab code was prepared to solve the above example in IGA context. The stiffness matrix for a single element is calculated and assembled into the global matrix following these steps: 
\begin{itemize}
\item Set $K=0$
\item loop over number of elements
\item Loop over Gaussian points $\tilde{\xi}_i , \tilde{w}_i$
\begin{enumerate}
\item find the parametric coordinate $ \xi = \tilde{\phi}^e(\tilde{\xi}_j)$
\item compute the basis function $R_a^e (a=1, ... p+1)$ at point $\xi$
\item define vector $\vect{R}= [ R_1^e, R_2^e, ..., R_{p+1}^e]$
\item compute the basis function derivatives $R_{a,\xi}^e (a=1, ... p+1)$ at point $\xi$
\item define vector $\vect{R}_\xi = [ R_{1,\xi}^e, R_{2,\xi}^e, ..., R_{p+1,\xi}^e]$
\item compute $|J_\xi| = \| \vect{R}_\xi \vect{P} \|$
\item compute $|J_{\tilde{\xi} }| = 0.5 (\xi_{i+1}-\xi_{i})$
\item compute shape function derivatives $\vect{R}_x = J_\xi^{-1} \vect{R}_{\xi}^T$
\item Set $\vect{k}^e = \vect{k}^e + \vect{R}_x \vect{R}_x^T|J_\xi||J_{\tilde{\xi} }| \tilde{w}_j  -k^2 \vect{R} \vect{R}^T|J_\xi||J_{\tilde{\xi} }| \tilde{w}_j   $
\item End of both loops
\end{enumerate}
\end{itemize}
then, the Neumann boundary condition is applied
and the resulting linear system is solved to find the estimated solution in IGA. 
we plot the approximate solution $ u_h$ for $k=40$ as well as the exact solution in Fig.\ref{fig:IGAFEM}. The density of discretization for IGA approximations is $ n_{\lambda} = 10$.  
As expected, IGA ($p=1$) also suffers from pollution error. B-Spline functions generated with $ p=0 $ and $ p=1$ orders will result in the same piecewise constant and linear functions as standard FEM shape functions. The differences are found for cases using higher order shape functions. The $ p^{th} $ order B-Spline function has $ p-1 $ continuous derivatives across the element boundaries. Each B-Spline basis is point-wise non-negative over the entire domain as a result, all of the entries of the mass matrix will be positive. The support of a B-Spline function of order $p$ is always $ p+1$ knot spans. Therefore, higher order B-Spline function has support over much larger portion of the domain when compared to classical FEM. The total number of functions that any given function shares support with (including itself) is $ 2p+1 $ regardless of whether a FEM basis or a B-Splines is used. Hence, using higher order B-Spline basis functions will provide improved support compared to standard FEM without increasing the number of required shape functions.
Increasing the order of Lagrangian polynomials will increase the amplitude of oscillations in FEM; this problem is eliminated in IGA as a result of non-negativity and non-interpolatory nature of B-Splines shape functions.

\begin{figure}[!hbt]
\centering
\includegraphics[width=1\textwidth]{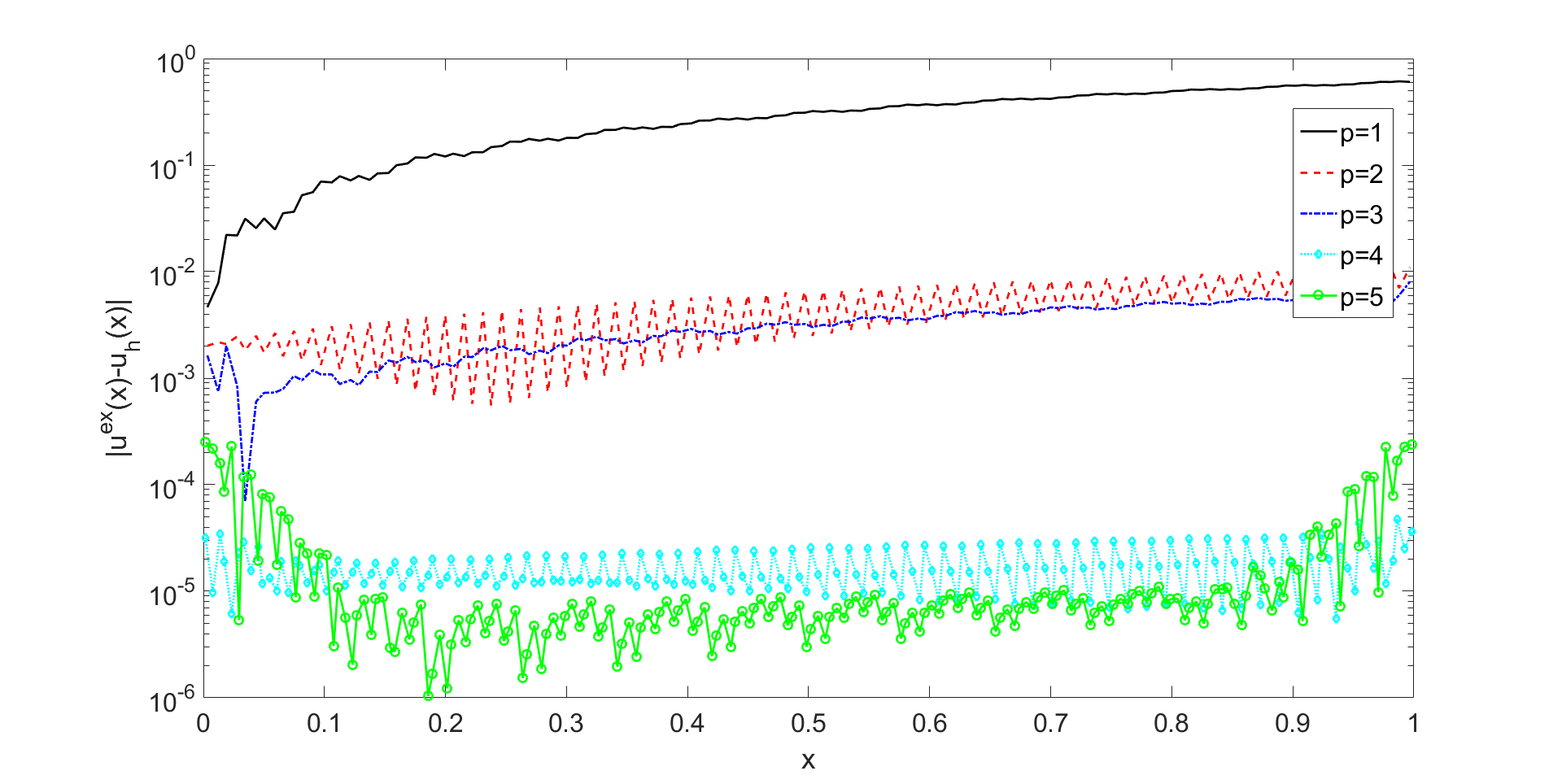}
\caption{Absolute error  $ |u^{\textrm{ex}} -u_h|$ for $k=40$, $n_{\lambda}=10$ and various approximation orders $p=1,..., 5$.}
\label{fig:1DerrP}
\end{figure}

\begin{figure}[!hbt]
\centering
\includegraphics[width=1\textwidth]{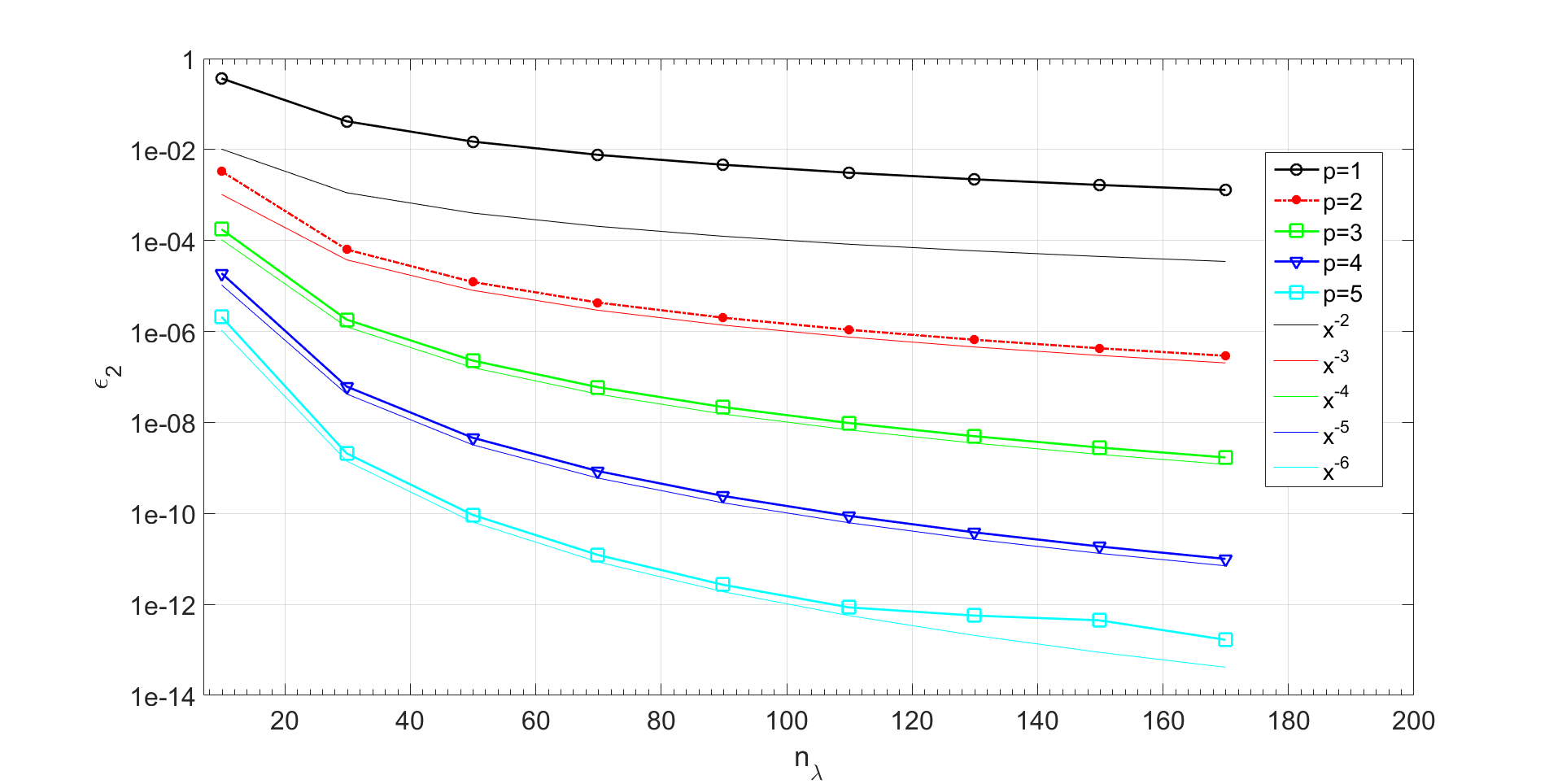}
\caption{Evolution of the error $\epsilon_2$ vs. the discretization density $n_{\lambda}$ for $k=40$ and $p=1,...,5$
(the slopes of the functions $x^{-p}$ are plotted for comparison).}
\label{fig:1Dconv}
\end{figure}

\begin{figure}[!hbt]
\centering
\includegraphics[width=1\textwidth]{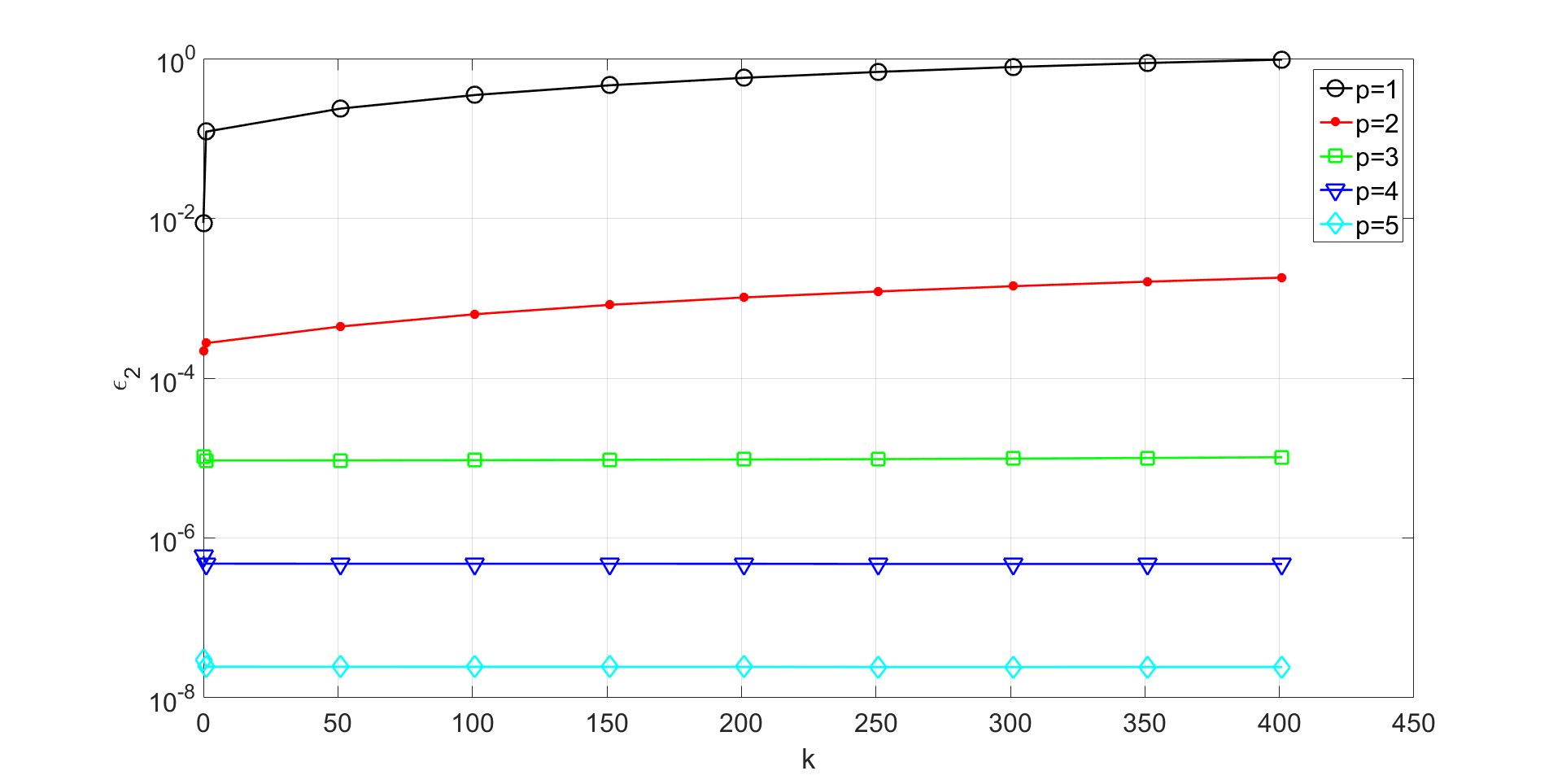}
\caption{Evolution of the error $\epsilon_2$ vs. the wavenumber $k$, for $n_{\lambda}=20$ and various
approximation orders $p=1,...,5$.}
\label{fig:1DL2K}
\end{figure}

Next we investigate the effect of increasing the basis function order on the accuracy of the IGA approximation again for discretization density $n_{\lambda} = 10$. As expected increasing the order $p$  reduces the error drastically as shown in Fig. \ref{fig:1DerrP}. There is no visible pollution error in the IGA approximation if we keep all the parameters unchanged and increase the order of the basis function.
One can reduce the error further by increasing the discretization density $n_{\lambda}$. For the numerical solution $f^{\textrm{calc}}(\vect{x}), \vect{x} \in \Omega_b$, we define the $L_2$-error (in dimension $d$) as
\begin{equation}
\epsilon_2  = \Big\{ \int_{\Omega_b} | f^{\textrm{calc}} (\vect{x}) - f^{\textrm{ex}}(\vect{x})|^2 d\vect{x} \Big\}^{1/2}. 
\label{L2error}
\end{equation}
The evolution of the $L_2$-error vs. the discretization density $n_{\lambda}$ is reported  in Fig. \ref{fig:1Dconv} for
the wavenumber $k=40$.  For completeness, we also plot the curves relative to the function $x^{-p}$, for $x=n_{\lambda}$.
We can see that the error curves fits very well the polynomial curves. This lets to think that the pollution error is extremely
weak, most particularly when $p$ is large enough ($p\geq3$) even for a small density $n_{\lambda}$.
Next we present the evolution of the $L_2$-error with the wavenumber $k$ in Fig. \ref{fig:1DL2K}
for a fixed density $n_{\lambda}=20$, where for
 $p=3$ and higher the pollution error is not visible and the error remains constant even for extremely high frequencies. 
\section{Two-dimensional examples}
\label{sec:2D}
In this section, we evaluate the performance of IGA in solving two-dimensional acoustic problems. First,
 we consider a two-dimensional duct problem with rigid walls \cite{Huttunen2008}. Next, we 
 propose two examples of cylindrical disk scattering problems where we obtain, first the scattered field of a particular mode,
   and then the scattering of the disk subject to an incident plane wave. By separating the truncation error from
   the numerical basis approximation, we fairly investigate the performance of IGA in solving
   the two-dimensional sound-hard exterior scattering problem and numerically analyze the related pollution error.
\subsection{The duct problem}
\begin{figure}[!htb]
\centering
\includegraphics[width=0.8\textwidth]{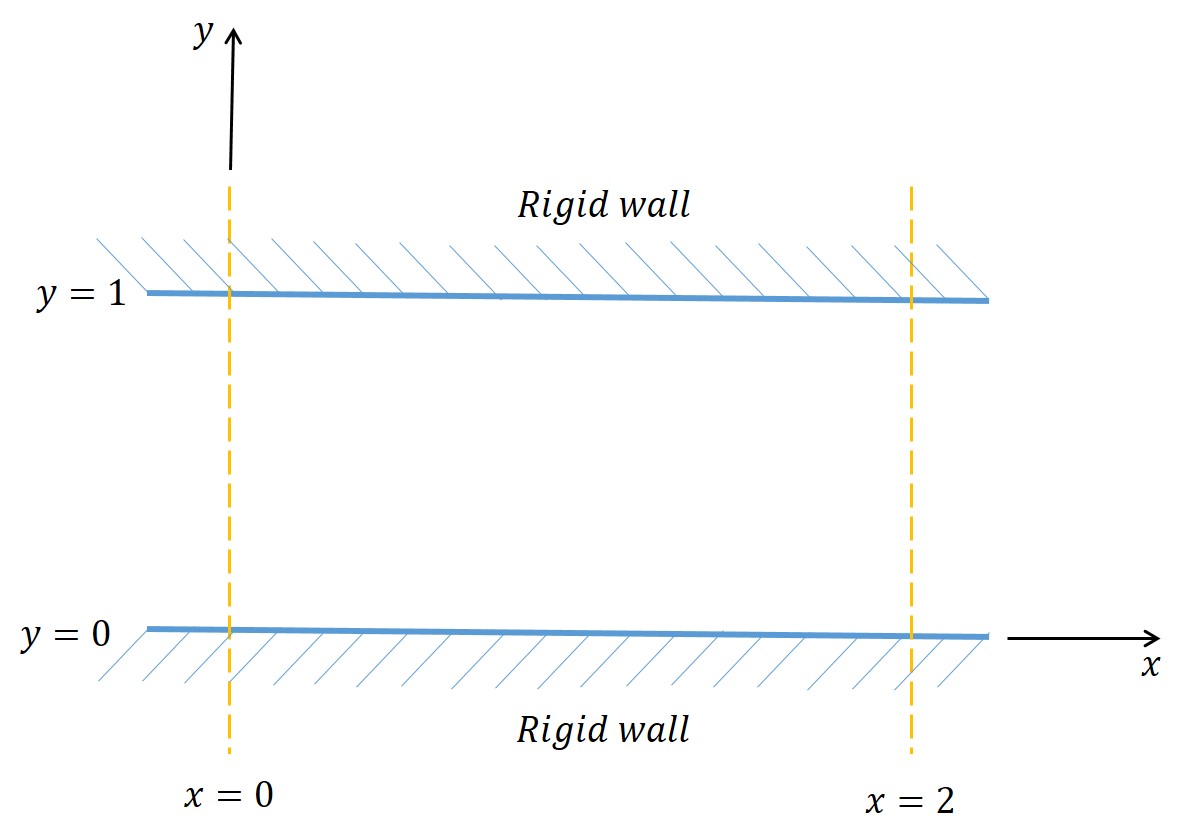}
\caption{The duct domain.}
\label{fig:Duct}
\end{figure}

\begin{figure}[!htb]
    \centering
        \begin{subfigure}[b]{0.48\textwidth}
        \includegraphics[width=\textwidth]{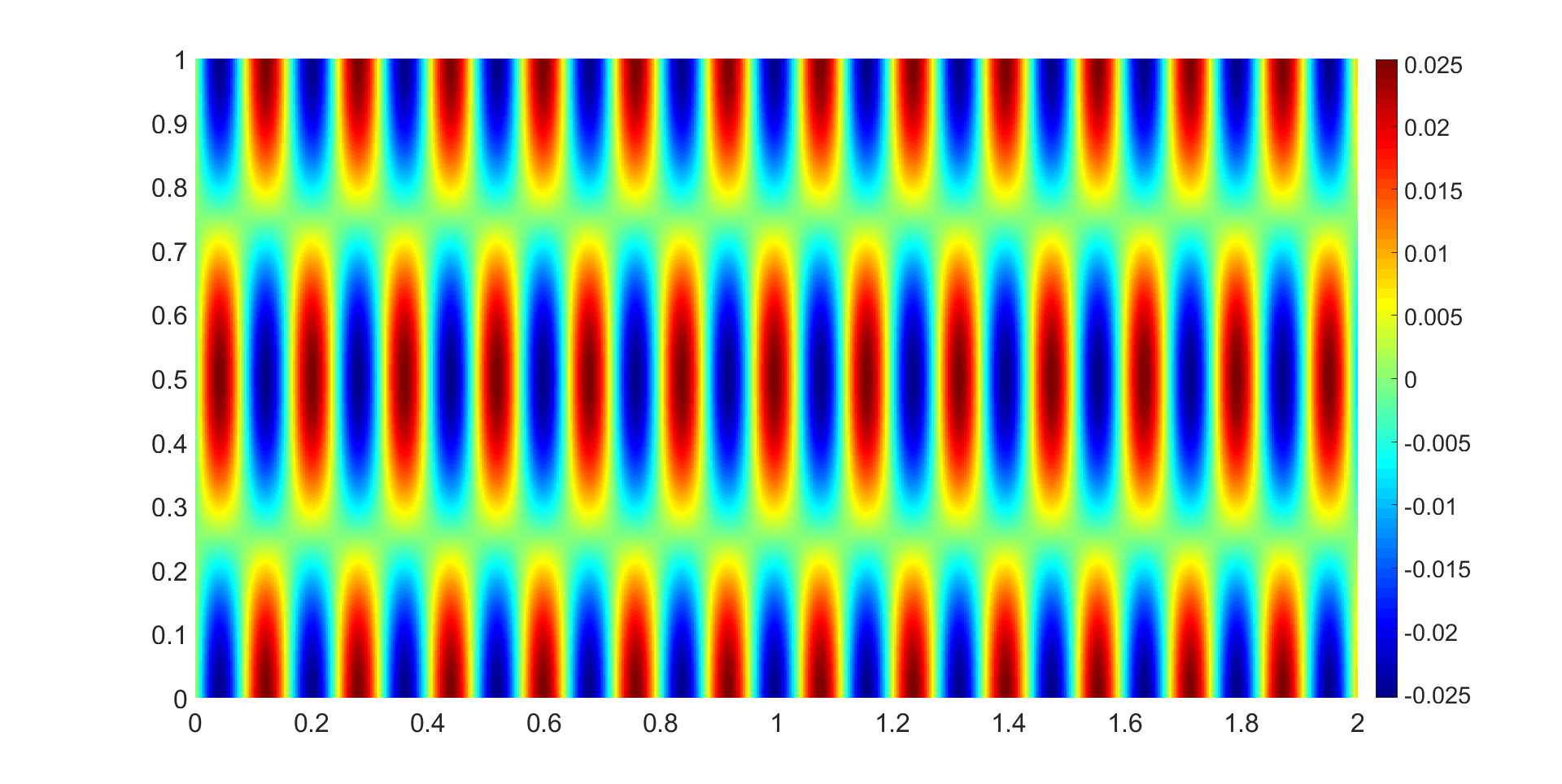}
        \caption{$\Re(u^{\textrm{ex}})$}
        \label{fig:Duct_real_exact}
    \end{subfigure} 
    \begin{subfigure}[b]{0.48\textwidth}
        \includegraphics[width=\textwidth]{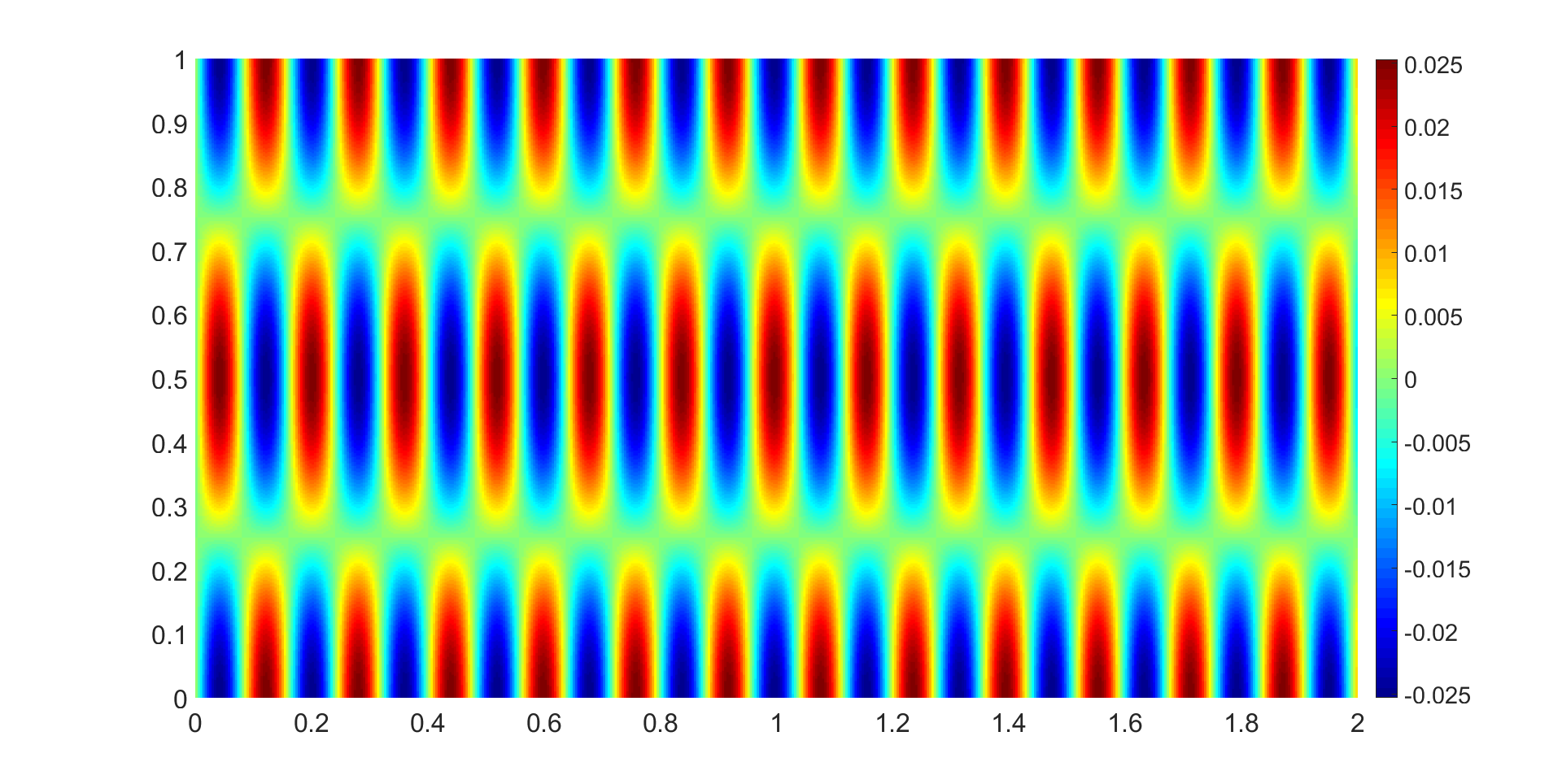}
        \caption{$\Re(u_{h})$}
        \label{fig:Duct_real_num}
    \end{subfigure}
   \caption{Comparing (a) the real parts of the exact solution $u^{\textrm{ex}}$ 
  and (b) the numerical solution $u_{h}$  for $k = 40$, $m=2$, $p=3$, and $n_{\lambda } =10$.}
\label{fig:Duct_real}
\end{figure}

\begin{figure}[!htb]
\centering
\includegraphics[width=0.8\textwidth]{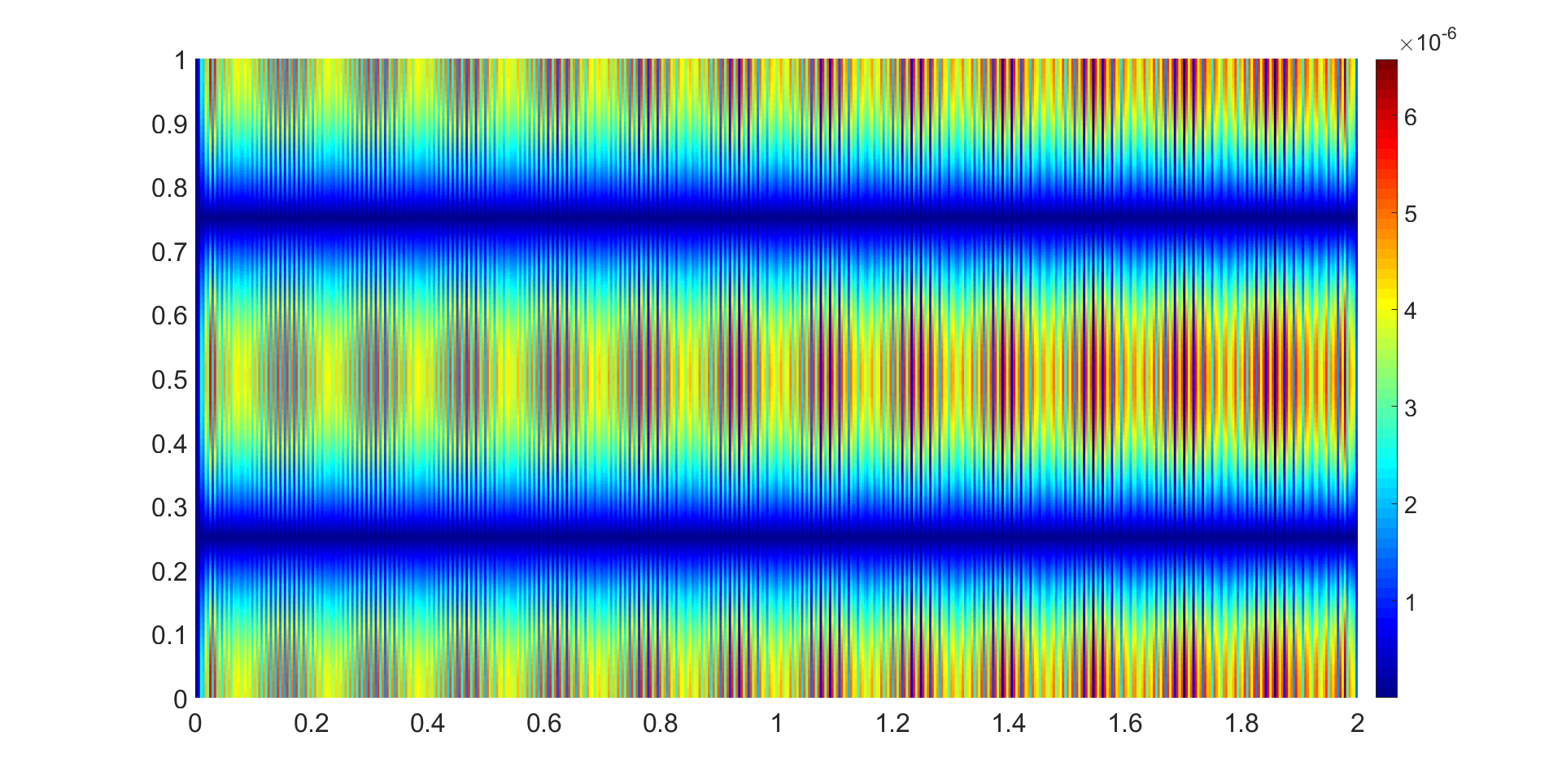}
\caption{Absolute error $|u^{\textrm{ex}}-u_h|$ for $k = 40$, $m=2$, $p=3$ and $n_{\lambda } =10$. }
\label{fig:Duct_error}
\end{figure}

\begin{figure}[!htb]
\centering
\includegraphics[width=\textwidth]{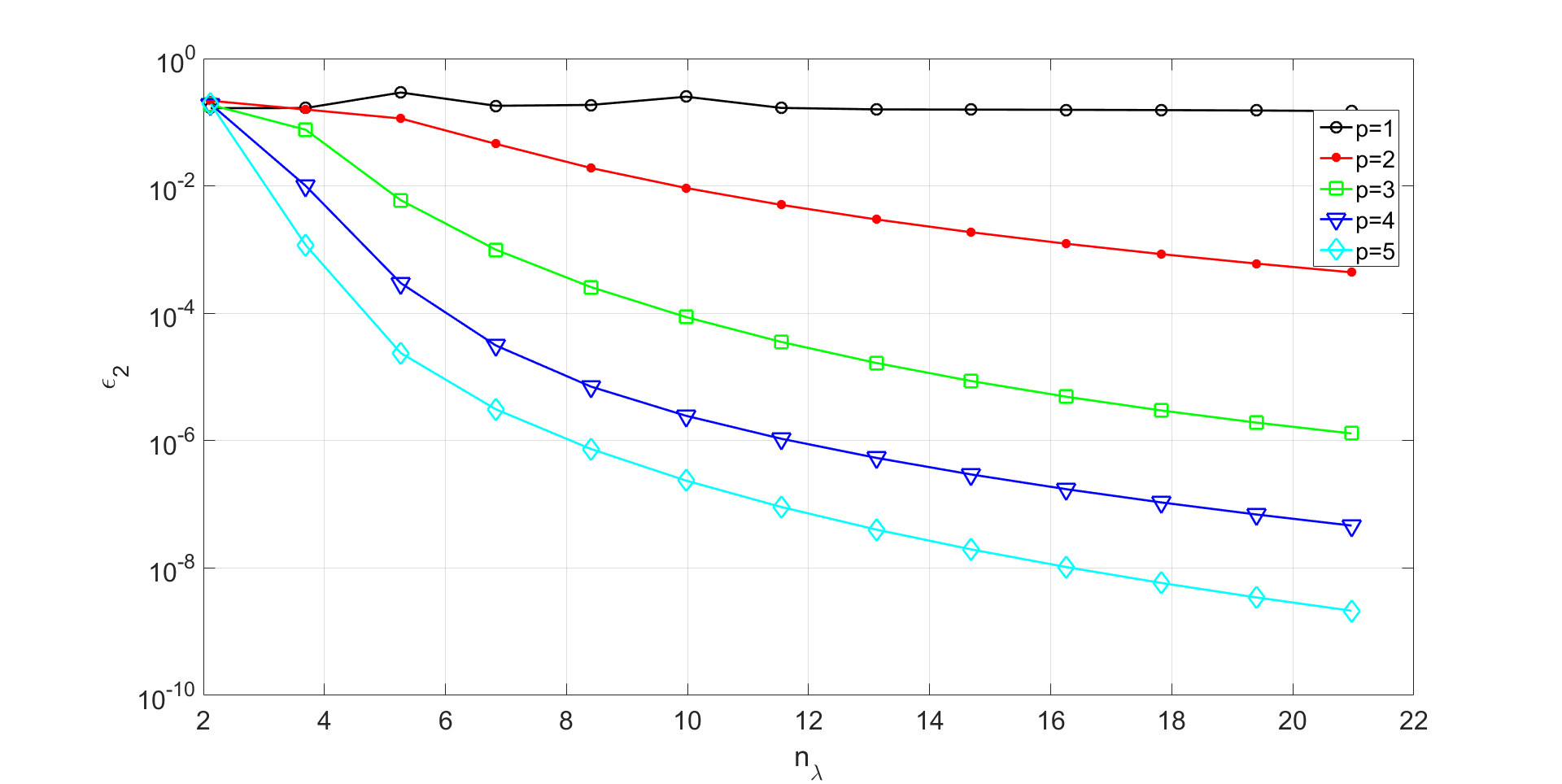}
\caption{Evolution of the error $\epsilon_2$ vs. the discretization density $n_{\lambda}$, for $k=40$, $m=2$
and $p=1,...,5$.}
\label{fig:Duct_conv}
\end{figure}

\begin{figure}[!htb]
\centering
\includegraphics[width=\textwidth]{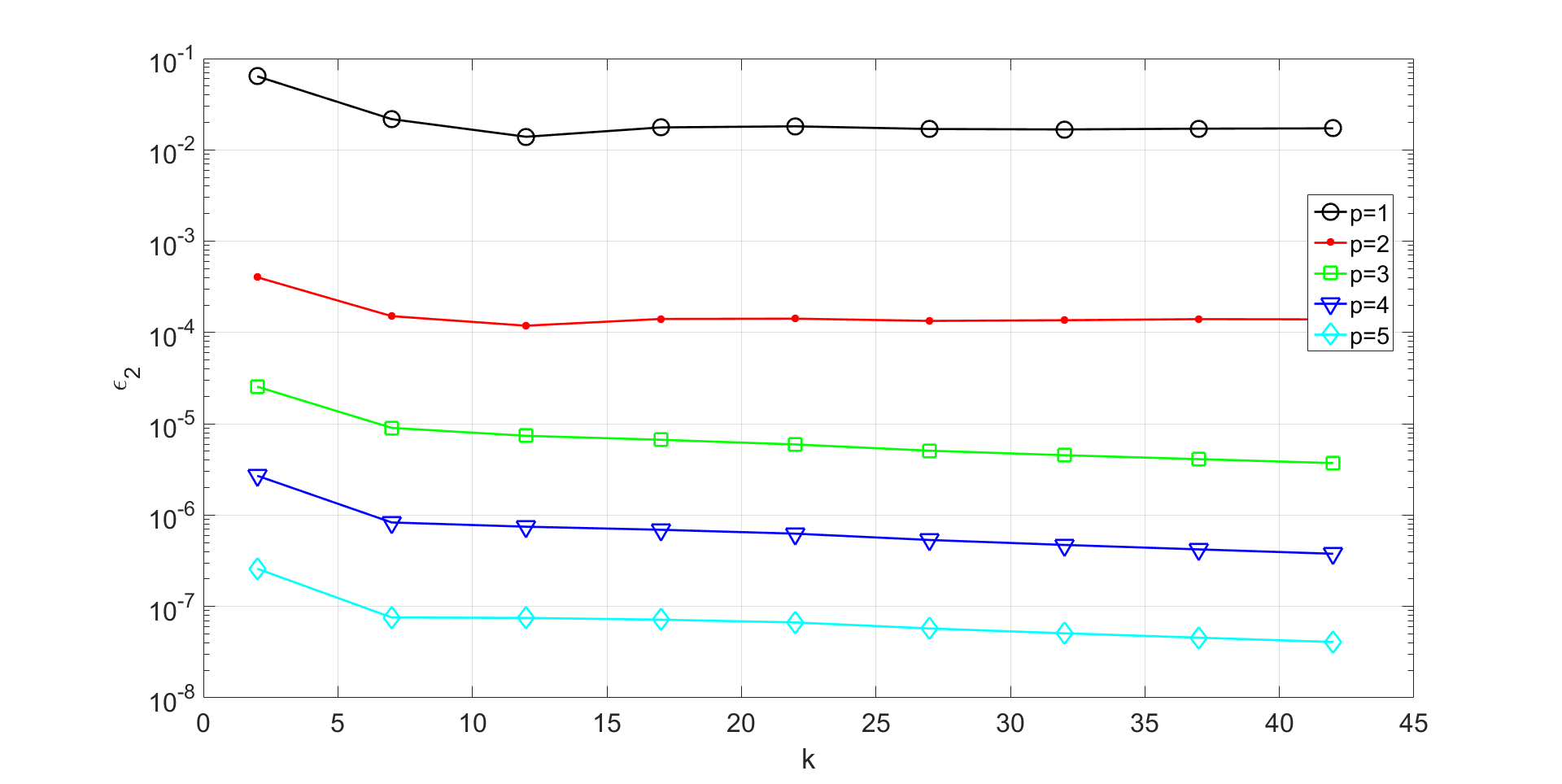}
\caption{Evolution of the error $\epsilon_2$ vs. the wavenumber $k$, for $m=2$, $n_{\lambda}=10$
and $p=1,...,5$.}
\label{fig:Duct_convK}
\end{figure}

The duct model is shown in Fig. \ref{fig:Duct} where $\Omega_{b}= [0,2] \times [0,1]$. We denote the outward boundary unit normal with $ v $ and assume that the lower and upper walls are rigid. 
We solve the Helmholtz problem stated in Eq. \ref{eq:duct} for the acoustic pressure $u$
\begin{equation}
\begin{aligned}
&\Delta u + k^2 u = 0, \quad &\text{in} \ \Omega_{b}, \\
&\frac{\partial u}{\partial v} = \cos (m \pi y), \quad &\text{on} \ x=0, \\
&\frac{\partial u}{\partial v} +iku = 0, \quad &\text{on} \ x=2,\\
&\frac{\partial u}{\partial v} = 0, \quad &\text{on} \ y=0, 1,
\label{eq:duct}
\end{aligned}
\end{equation}
where $ m \in \mathbb{N} $ is the mode number. An inhomogeneous boundary condition is applied on the inlet boundary 
($ x = 0$) and an absorbing (and transparent for $m=0$) boundary condition is set on the outlet boundary ($ x = 2 $). Since the  boundaries at $ y = 0, 1$ are assumed to be perfectly rigid, the normal derivative of the  acoustic pressure vanishes on these boundaries. The exact solution of  problem (\ref{eq:duct}) with ABC is as follows
\begin{equation}
u^{\textrm{ex}} (x,y) = \cos (m \pi y) (A_1 e^{-i k_x x} + A_1 e^{i k_x x} ),
\end{equation}
where $ k_x = \sqrt{k^2-(m \pi)^2} $ and the coefficients $ A_1 $ and $ A_2 $ are obtained from 
\[
\begin{pmatrix}
    k_x       & -k_x \\
    (k-k_x)e^{-2ik_x} &(k+k_x)e^{2ik_x}\\
\end{pmatrix}
\begin{pmatrix}
   A_1 \\
    A_2 \\
\end{pmatrix}
=
\begin{pmatrix}
   1 \\
   0 \\
\end{pmatrix}.
\]
Solving this $2\times 2$ linear system, we are able to get the expression of the solution to the duct problem with ABC. This is a very
interesting point since we therefore can only focus the analysis on the finite-dimensional approximation without
including the (continuous) truncation error. 
The cut-off frequency is $m_{\text{cut-off}} = k/\pi $. If the mode is such that $ m \leq m_{\text{cut-off}} $, the solution is representing propagating modes and if $ m > m_{\text{cut-off}} $ the solution corresponds to evanescent modes. The real parts of the  exact and estimated IGA solutions are presented in Fig. \ref{fig:Duct_real_exact} and
\ref{fig:Duct_real_num}, respectively,
 for $ k = 40$,
$ m =2 $ (propagative mode),
$ p =3$, and $ n_{\lambda} = 10 $. The corresponding
 absolute error $ |u^{\textrm{ex}}-u_h| $ is plotted in Fig. \ref{fig:Duct_error}, the maximal value being of the order
 of $10^{-6}$.
For the propagative mode $m=2$, the evolution of the $L_2$-error with respect to
the  discretization density $n_{\lambda}$ is shown in Fig. \ref{fig:Duct_conv} for $k=40$,
and according to the wave number $k$ in Fig. \ref{fig:Duct_convK}, where $n_{\lambda} = 10$.
 We can see that the error is very low, decreasing strongly with $p$. In addition, the error does not seem to depend on $k$
 as observed in Fig. \ref{fig:Duct_convK}, meaning that the pollution error is negligible.

\newpage
\subsection{The sound-hard circular cylinder problem}
To further study the performance of IGA, we analyze the scattering by a 2D circular cylinder $\Omega^- = \mathcal{D}_0$ of radius $ R_0 $ centered at the origin, with boundary $ \Gamma = \mathcal{C}_0 $ (circle of radius $R_{0}$).
The scatterer is surrounded by an outer fictitious circular boundary $ \Sigma = \mathcal{C}_1 $, again centered at the
origin and with radius $ R_1 > R_0 $. Hence, the computational domain $ \Omega_b$ is the annulus
bounded between the inner $\mathcal{C}_0 $ and outer $\mathcal{C}_1 $ boundaries.
On the exterior boundary, we set the symmetrical second-order Bayliss-Turkel ABC \cite{AntoineBarucqBendali}� given by
\begin{equation}
\label{BGTABC}
\partial_{\mathbf{n}_{\Sigma}}u+(-ik+\frac{\kappa}{2}-\frac{\kappa^2}{8(\kappa-ik)})u
-\frac{1}{2(\kappa-ik)}\partial_s^2u=0, \ \ \text{on } \Sigma,
\end{equation}
 setting $\mathbf{n}_{\Sigma}$ at the outwardly directed unit normal to $\Sigma$, 
 $\partial_{\mathbf{n}_{\Sigma}}:=\partial_{r}$ the normal derivative,
 $\kappa= 1/R_{1}$
 is the curvature of $\Sigma := C_{1}$ and $\partial_s^2:=R_{1}^{-2}\partial_{\phi}^2$ is the second-order curvilinear derivative
 on $C_{1}$. In the above expressions, the polar coordinate system is denoted by $(r,\phi)$.
\subsubsection{Scattering by a sound-hard circular cylinder : mode-by-mode analysis}
We consider an incident wave with a fixed mode   $m$ which is described as
\begin{equation}
u_m^{\textrm{inc}}(\vect{x}) = J_m(kr)e^{im\phi}, \quad m \in \mathbb{Z},
\end{equation} 
where $J_{m}$ is the $m$-th order Bessel's function.
The exact exterior modal solution $u_m^{\textrm{ex}}$ to the truncated scattering problem of the inner sound-hard circular
cylinder $\mathcal{C}_0$ in polar coordinate $ (r,\phi) $ is given as 
\begin{equation}
u_m^{\textrm{ex}}(\vect{x}) = \left(a_m H_m^{(1)}(kr)+b_m H_m^{(2)}(kr) \right)e^{im\phi}, \quad r\geq R_0, \quad m\in\mathbb{Z}.
\label{eq:Disk_exact}
\end{equation}
The functions $H_m^{(1)}$ (respectively $H_m^{(2)}$) is the first-kind (respectively second-kind)
Hankel function of order $m$.
The Neumann boundary condition is applied on $\mathcal{C}_0$ and the ABC on $\mathcal{C}_1$,
 resulting in the following linear system of equations to obtain the two unknown coefficients $a_m$ and $b_m$
\begin{equation}
a_m = -\frac{A^{m}_{22}J_m'(kR_0)}{D_{m}}, \quad b_m = -\frac{A^{m}_{21}J_m'(kR_0)}{D_{m}},
\end{equation}
where $ D_{m} = A^{m}_{11} A^{m}_{22} - A^{m}_{21} A^{m}_{12} $ and
\[ \begin{cases}
    A^{m}_{11} = H_m'^{(1)}(kR_0), \quad  &A^{m}_{21} = kH_m'^{(1)}(kR_1)-\mathcal{B}_m H_m^{(1)}(kR_1),\\
    A^{m}_{12} = H_m'^{(2)}(kR_0), \quad &A^{m}_{22} = kH_m'^{(2)}(kR_1)-\mathcal{B}_m H_m^{(2)}(kR_1),\\
    \displaystyle
    \mathcal{B}_m = -\big(\alpha_m \frac{m^2}{R_1^2} + \beta_m \big), \quad & \displaystyle \alpha_m = -\frac{1}{2ik}\big(1+\frac{i}{kR_1} \big)^{-1},\\
\displaystyle   \beta_m = -ik + \frac{1}{2R_1}+\frac{1}{8iR_1(1+kR_1)}.
\end{cases}
\]
The notation $f':=\partial_{r}f$ designates the radial derivative of a given function $f(r)$.
Quite similarly to the duct problem, the modes $m\in \mathbb{N}$ such that $|m|<  kR_{0}$ are propagating. For $|m| > kR_{0}$, the modes are evanescent (and therefore are not visible in the far-field since they do not
propagate). A special case corresponds to $|m| \approx kR_{0}$, which is related to grazing modes that
are tangent to the scatterer (generating glancing rays). 

\begin{figure}[b]
    \centering
    \begin{subfigure}[b]{0.35\textwidth}
        \includegraphics[width=\textwidth]{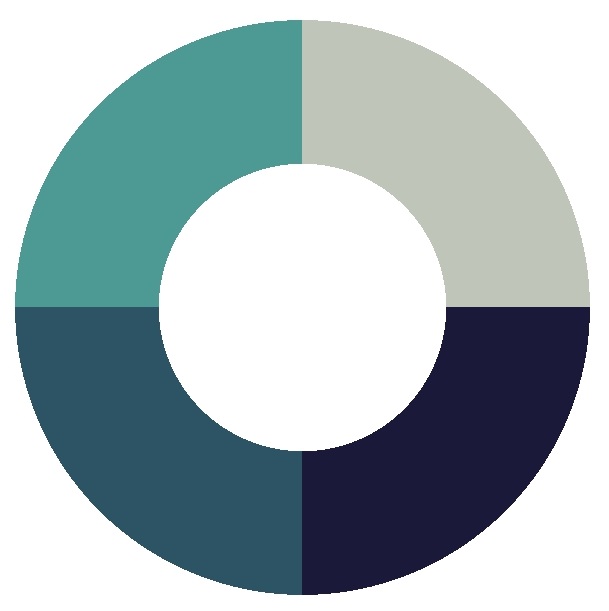}
        \label{fig:Disk_patches}
    \end{subfigure}
   \caption{Outline of the patches.
   }
\label{geometrydisk}
\end{figure}

The modal analysis is 
meaningful  since we can understand
the accuracy of the IGA approximations thanks to the spatial frequencies $m$ and the dimensionless
wavenumber $kR_{0}$. In addition, this also helps in clarifying what can be expected for the case of the scattering of a plane wave  by the disk since the exact solution is built as a modal series expansion of the elementary mode solutions. We fix $R_0 =1$ and $R_1 =2$. The IGA model was generated using four identical patches as shown in Fig. \ref{geometrydisk}.
We report the real parts of $u_{m}^{\textrm{ex}}$ and $u_{h}$ 
on Fig. \ref{fig:Disc_real_num} where the wavenumber is $k=40$ and $m = 2$ (which is
a propagative mode since $m=2 \leq kR_{0}=40$).
For IGA, the order of the method is $p=3$ for a density of discretization points per wavelength $n_{\lambda}=10$. 
We clearly see that the wavefield is accurately computed since the absolute error $|u_{m}^{\textrm{ex}}-u_{h}|$
is below $10^{-5}$ as it can be deduced from Fig. \ref{fig:Disk_error}. 
\begin{figure}[!htb]
    \centering
    \begin{subfigure}[b]{0.49\textwidth}
        \includegraphics[width=\textwidth]{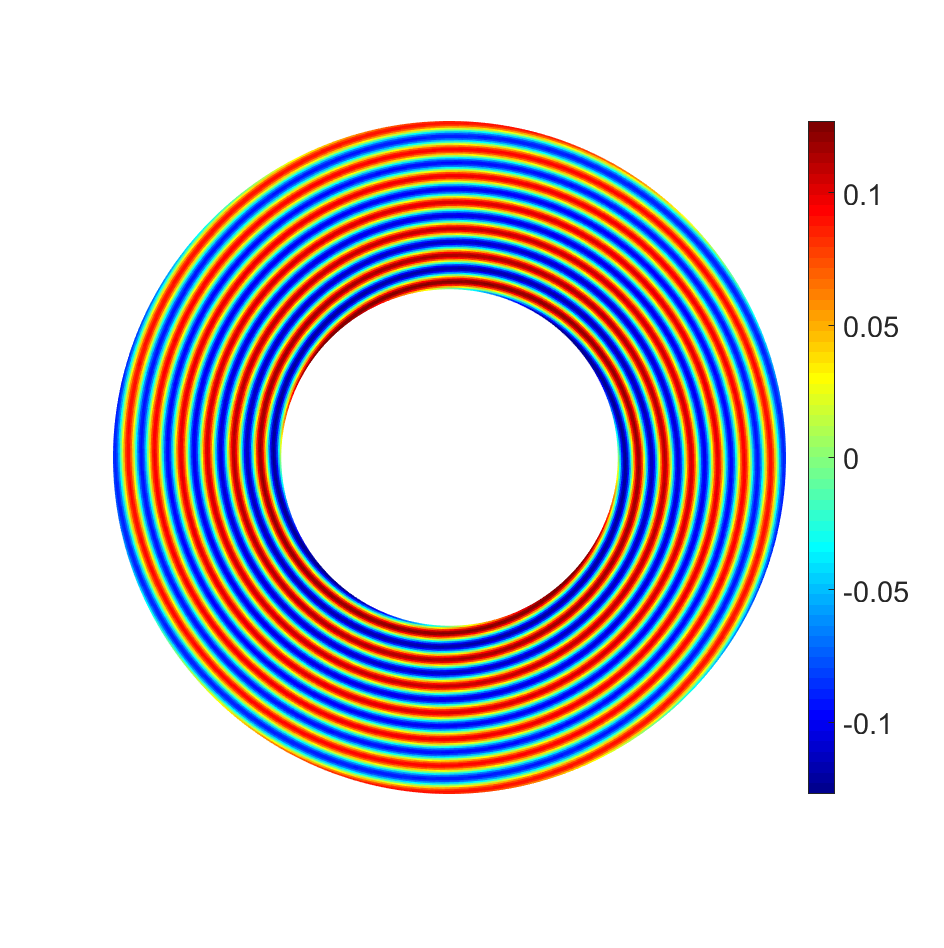}
        \caption{$\Re(u_{m}^{\textrm{ex}})$}
        \label{fig:Disc_real_exact}
    \end{subfigure}
    \begin{subfigure}[b]{0.49\textwidth}
        \includegraphics[width=\textwidth]{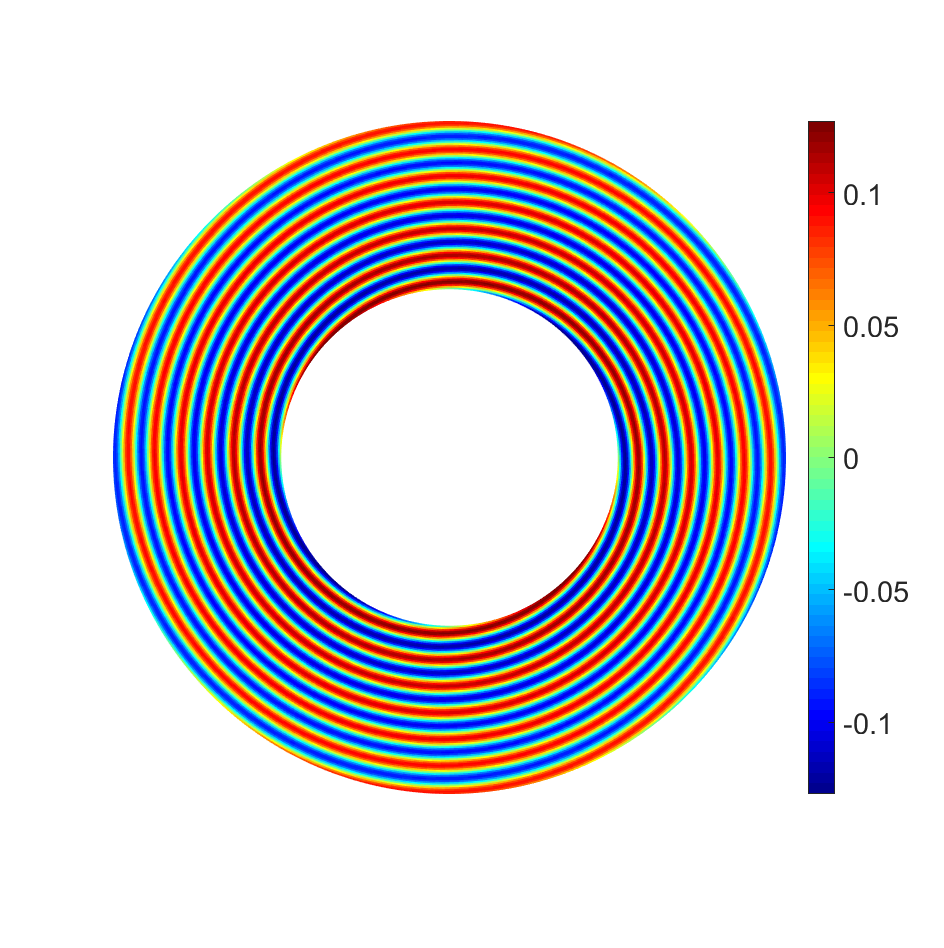}
        \caption{$\Re(u_{h})$}
        \label{fig:Disc_real_num}
    \end{subfigure} 
\caption{Comparing (a) the real parts of the
exact solution $u_{m}^{\textrm{ex}}$   and (b) the numerical IGA solution $u_{h}$
 for $k = 40$, $m=2$, $p=3$, and $n_{\lambda } =10$.}
\label{fig:Disc_real_num}
\end{figure}
\begin{figure}[!htb]
    \centering
     \begin{subfigure}[t]{0.5\textwidth}
        \includegraphics[width=\textwidth]{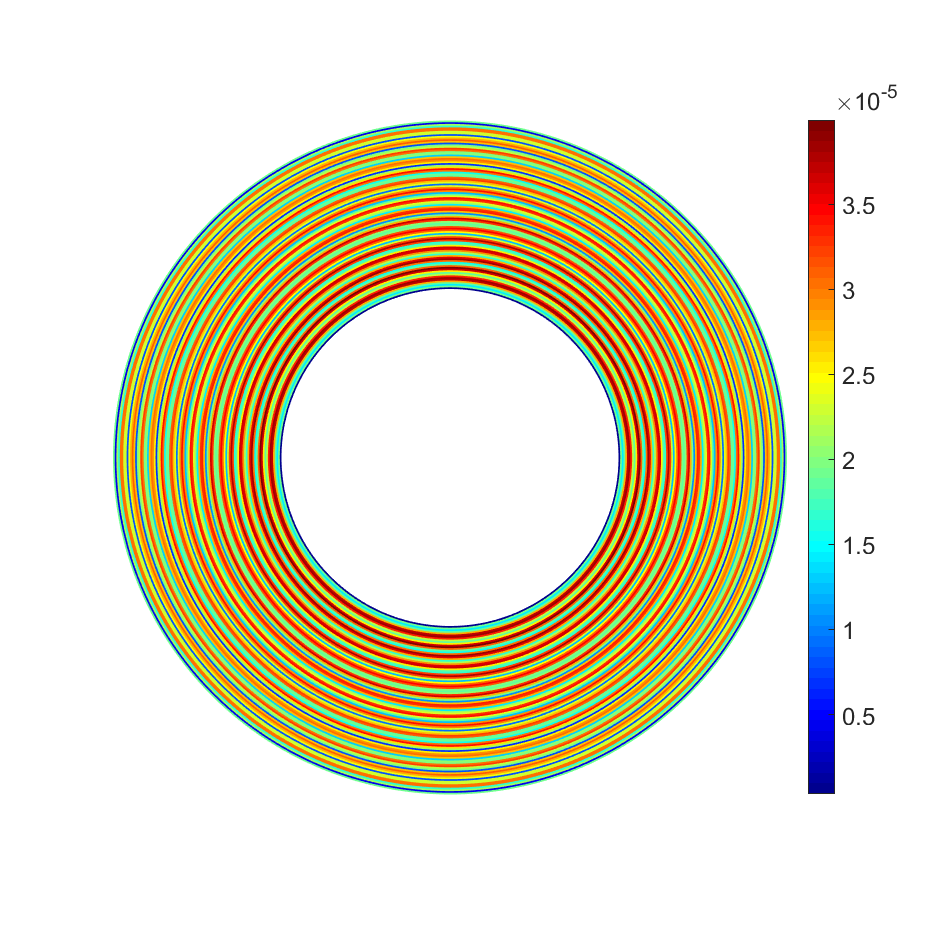}
    \end{subfigure} 
   \caption{Absolute error $|u_{m}^{\textrm{ex}}-u_h |$ for $k = 40$, $m=2$, $p=3$ and $n_{\lambda } =10$.}
\label{fig:Disk_error}
\end{figure}
The convergence graph presented in Fig. \ref{fig:Disk_conv} shows the effect of increasing the discretization density $n_{\lambda}$ 
for various orders $p$ of the underlying IGA basis function on the $L_2$-error $\epsilon_2$ for $k = 40$ and $m=2$.
We  see that the error curves follow the slopes of the curves $x^{-p}$ (for $x=n_{\lambda}$).
\begin{figure}[!htb]
\centering
\includegraphics[width=\textwidth]{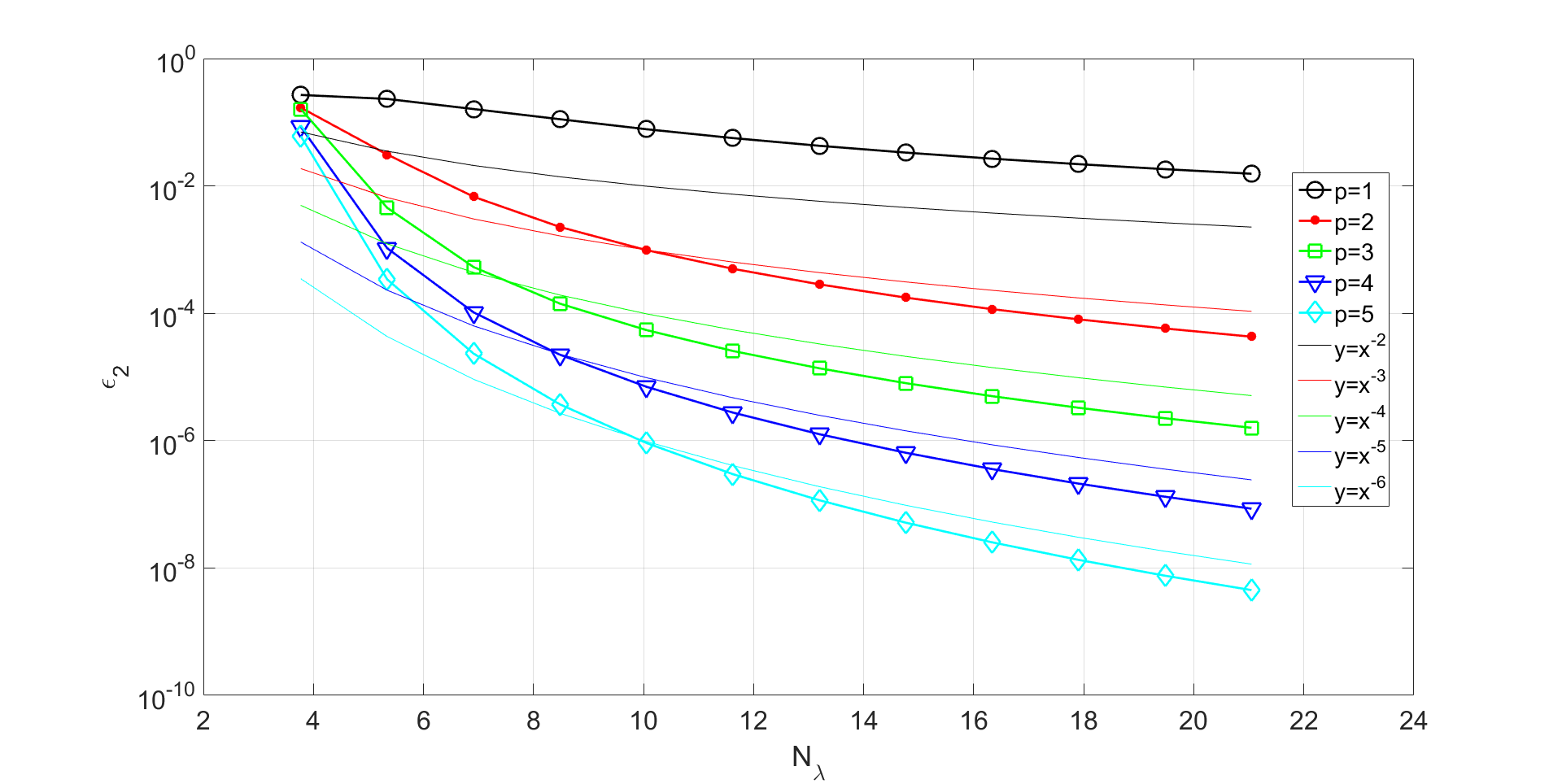}
\caption{Evolution of the error $\epsilon_2$ vs. the discretization density $n_{\lambda}$ for  $k=40$,
 $m=2$ and $p=1,...,5$.
}
\label{fig:Disk_conv}
\end{figure}
The evolution of the error $\epsilon_2$ with respect to the wavenumber $k$ is plotted on Fig. \ref{fig:Disk_convK},
for $m=2$, with various approximation orders $p$ of IGA and $n_{\lambda}=10$.  
We can observe that the accuracy does not really depend on $k$ for a fixed density when $p\geq2$.
%
\begin{figure}[!htb]
\centering
\includegraphics[width=\textwidth]{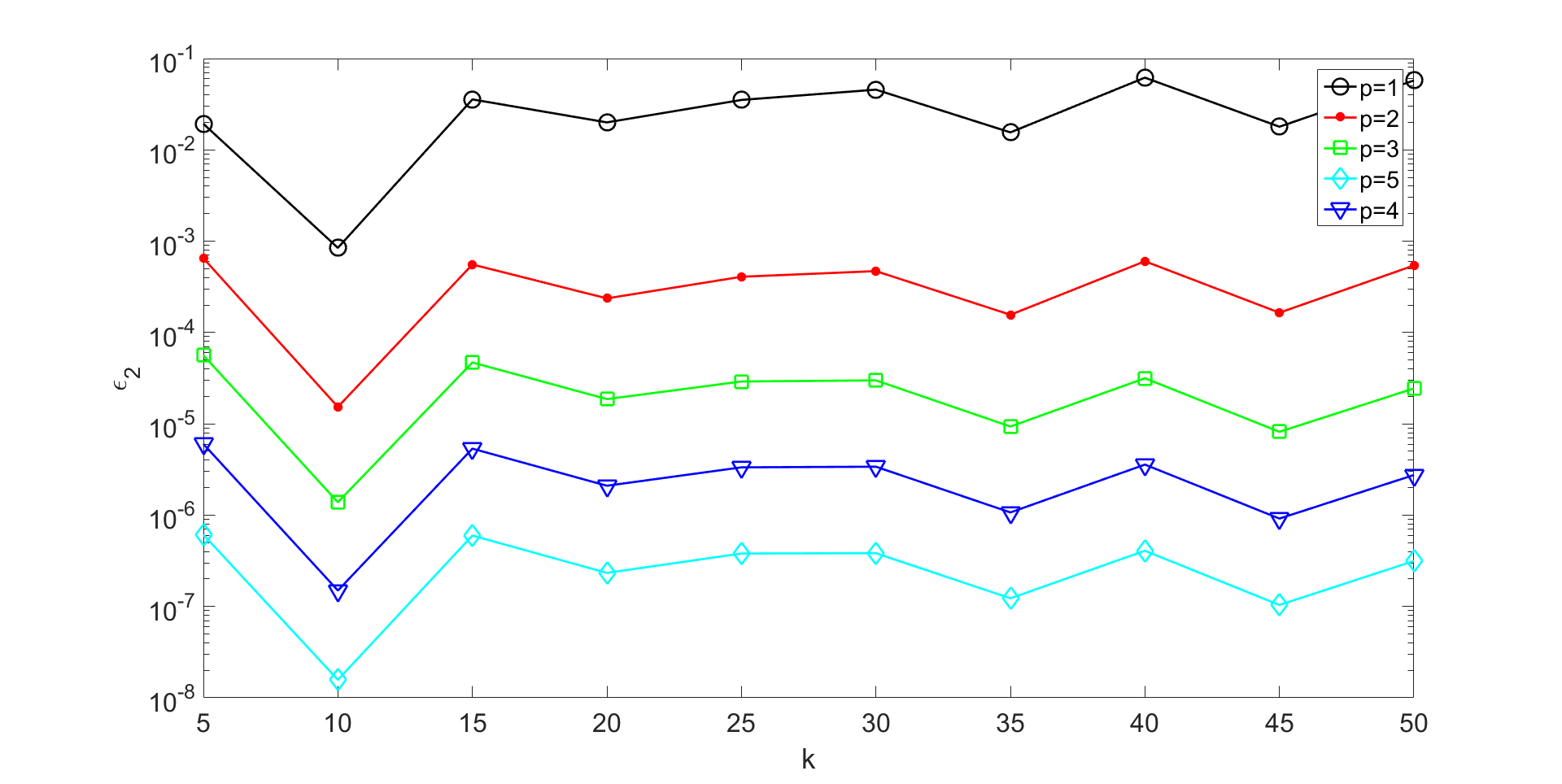}
\caption{Evolution of the error $\epsilon_2$ vs. the wavenumber $k$ for $m=2$, $n_{\lambda}=10$ and $p=1,...,5$.
}
\label{fig:Disk_convK}
\end{figure}
%
%
%
In fact, one question is to know how this error depends on $m$ for a fixed $k$. To analyze this point, we report on Fig.
\ref{fig:Disk_conv_mp} the evolution of the error $\epsilon_2$ thanks to $m$ and various values of $p$, for a fixed
framework $k=40$ and $n_{\lambda}=10$. We can remark that the error almost does not depend on $m$,
the value $m=2$ corresponding more or less to the worst situation. When $|m|\geq k R_{0}$, the error starts decreasing
strongly. Indeed, the solution is then evanescent and rather corresponds to the solution of a positive
definite problem (similar to a shifted laplacian). Such solutions are easier to compute with high accurary
 than for a propagative solution and do not participate to the  far-field. From these remarks,
 and since the solution of the scattering of a plane wave by a disk is computed as the superposition
 of these modes solutions, we can expect that similar graphs to Fig. \ref{fig:Disk_convK}
 could be obtained for the full plane wave problem (see subsection \ref{FullPlaneDisk}).
\begin{figure}[!htb]
\centering
\includegraphics[width=\textwidth]{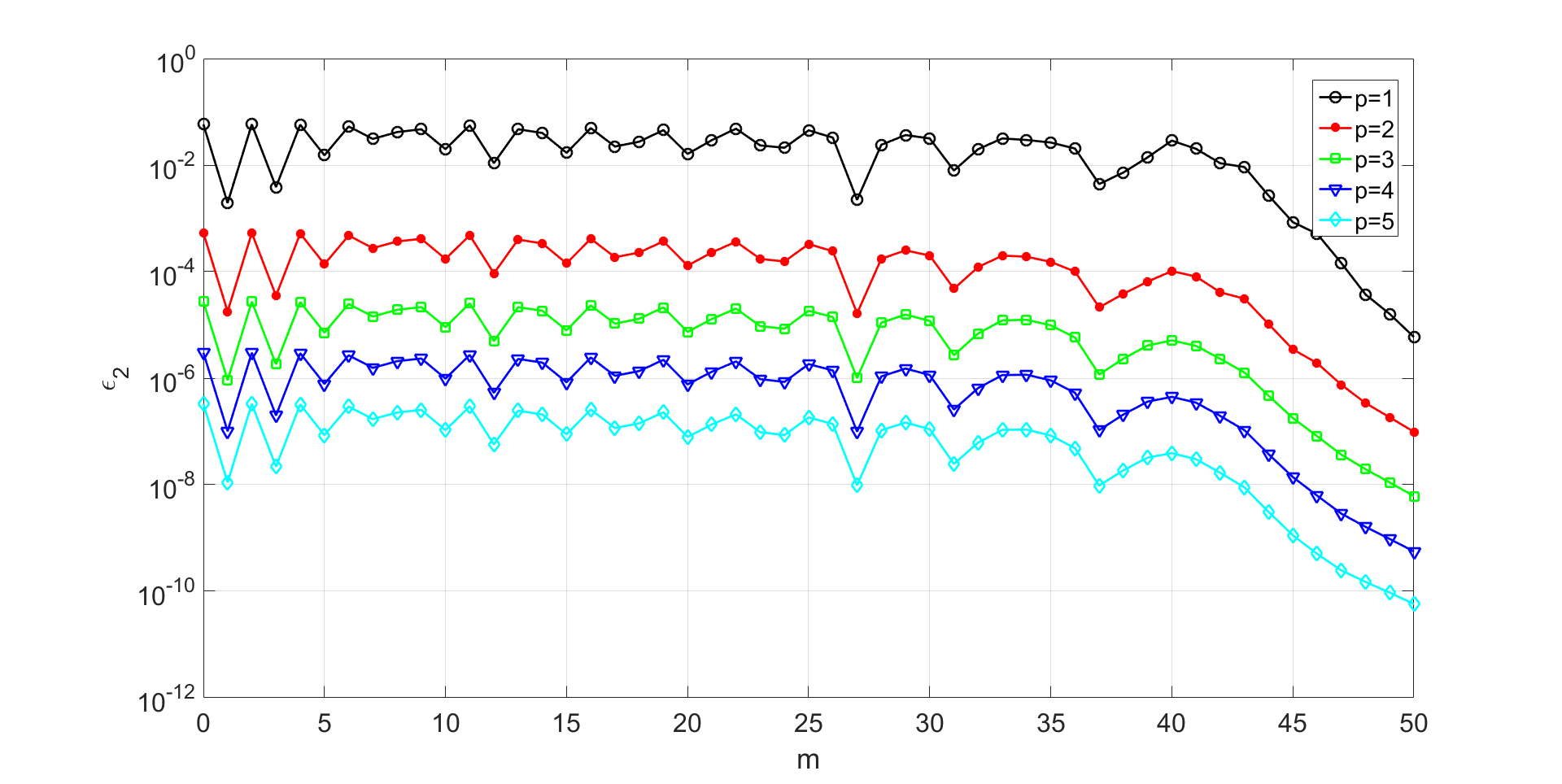}
\caption{Evolution of the error $\epsilon_2$ vs. the mode $m$, for $k = 40$, $n_{\lambda}=10$ and $p=1,...,5$.}
\label{fig:Disk_conv_mp}
\end{figure}
\subsubsection{Scattering by a sound-hard circular cylinder: plane wave scattering }\label{FullPlaneDisk}
\begin{figure}[!htb]
    \centering
     \begin{subfigure}[b]{0.49\textwidth}
        \includegraphics[width=\textwidth]{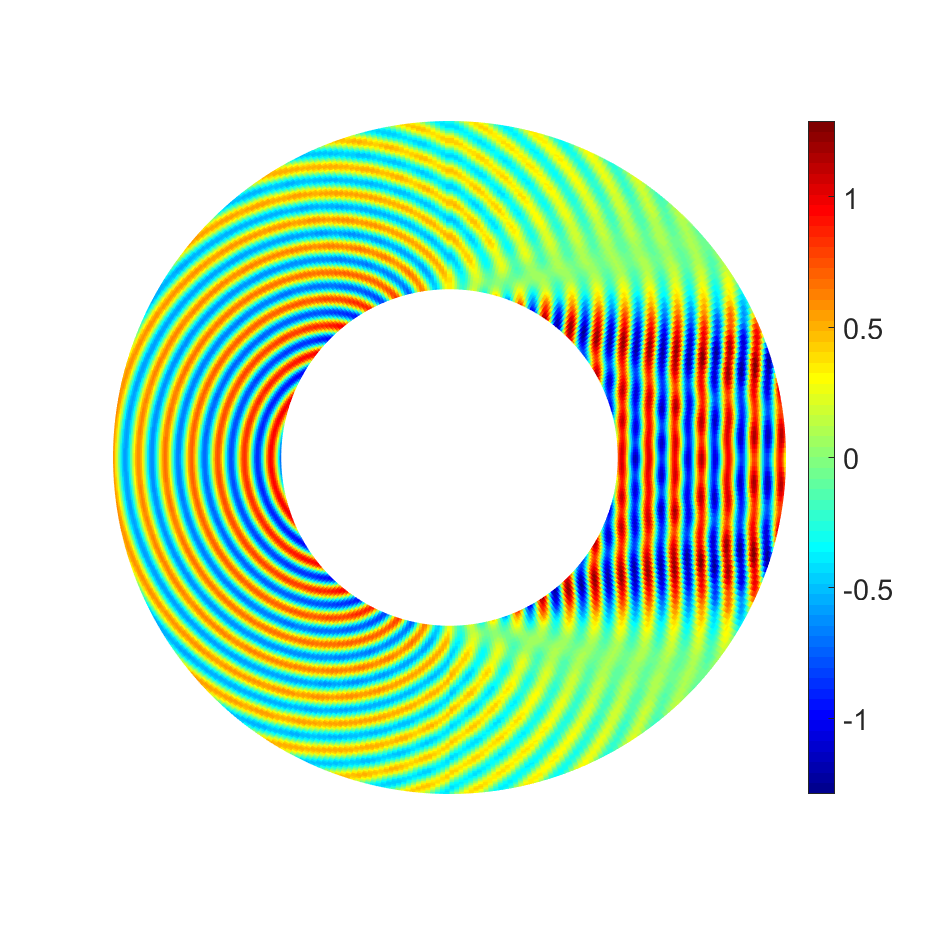}
        \caption{$\Re (u^{\textrm{ex}})$}
        \label{fig:Disk_imag_pw}
    \end{subfigure} 
    \begin{subfigure}[b]{0.49\textwidth}
        \includegraphics[width=\textwidth]{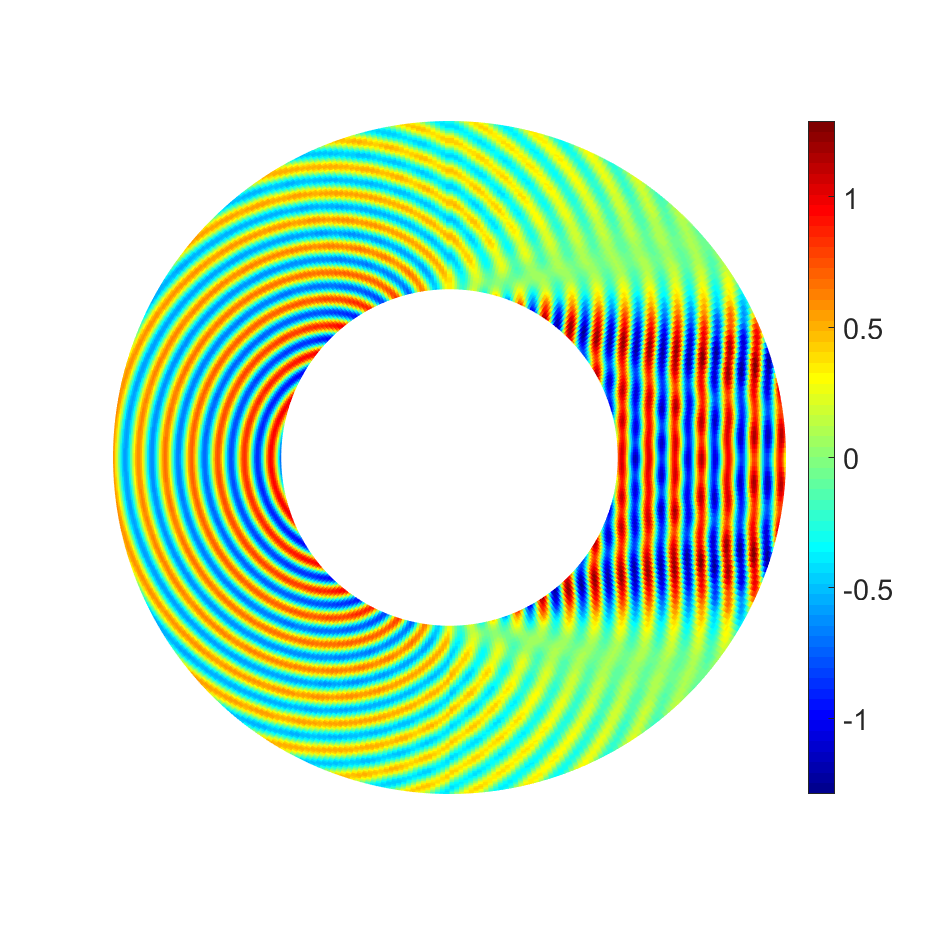}
        \caption{$\Re(u_h)$}
        \label{fig:Disk_real_pw}
    \end{subfigure}
    ~ 
   \caption{Comparing the real parts of (a) the
exact solution $u^{\textrm{ex}}$   and (b) the numerical IGA solution $u_{h}$ 
 for $k = 40$, $p=3$, and $n_{\lambda } =10$.}\label{fig:Disk_pw}
\end{figure}

\begin{figure}[!htb]
    \centering
     \begin{subfigure}[t]{0.49\textwidth}
        \includegraphics[width=\textwidth]{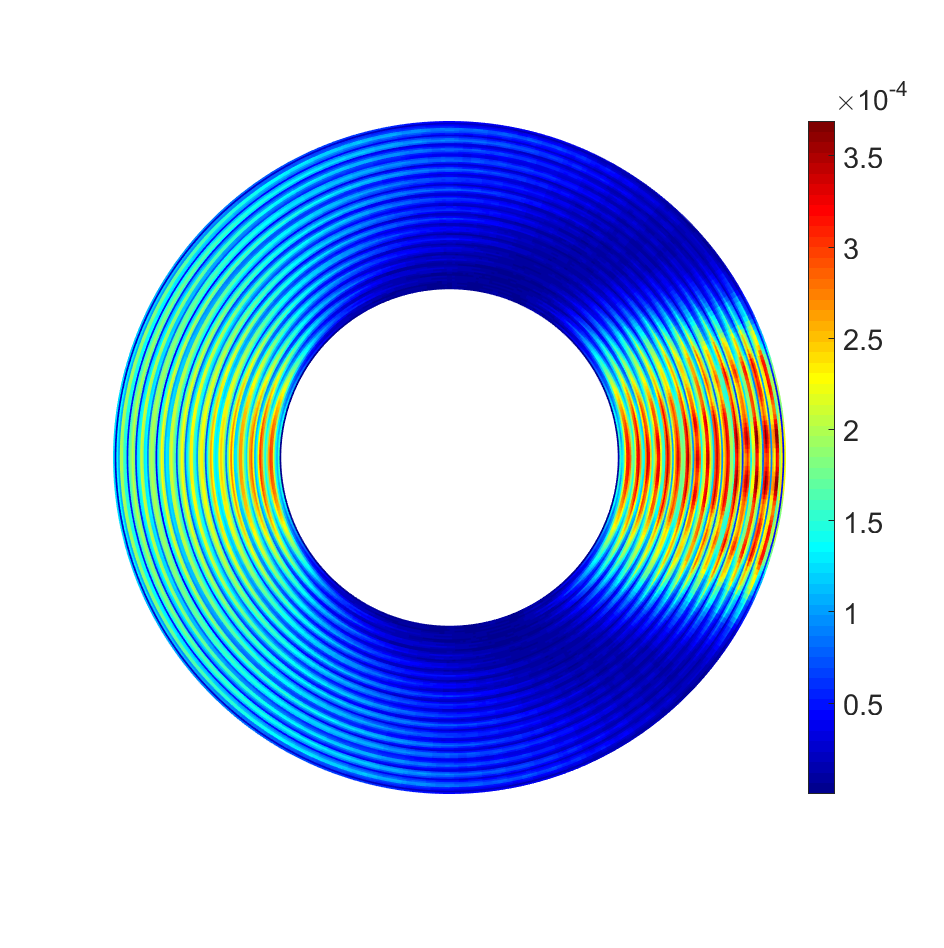}
            \caption{$n_{\lambda } =10$}
            \label{Fig17aa}
    \end{subfigure} 
        \begin{subfigure}[t]{0.49\textwidth}
        \includegraphics[width=\textwidth]{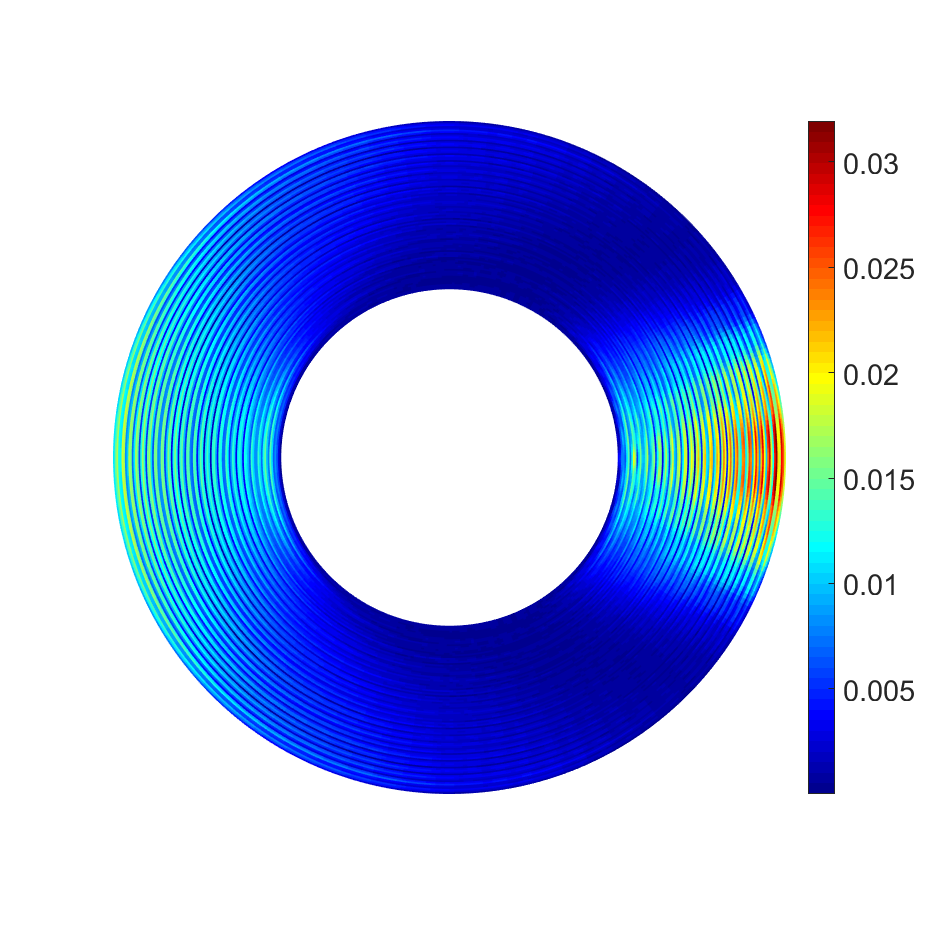}
               \caption{$n_{\lambda } =5$}
               \label{Fig17bb}
    \end{subfigure} 
   \caption{Absolute error $|u^{\textrm{ex}}-u_h |$ for $k = 40$ and $p=3$. The discretization density  is
    (a) $n_{\lambda } =10$ and (b) $n_{\lambda } =5$.}
\label{fig:Disk_pw_error}
\end{figure}

In the previous section, the  scattering problem of a sound-hard circular cylinder was found for a given harmonics $m$.
 The solution for the full plane wave incidence can be constructed by superposition
 as a Fourier series expansion of these harmonics. In this section, we consider an
   incident plane wave $ u^{\textrm{inc}} (\vect{x}) = e^{i k \vect{d} \cdot \vect{x}}$, where $ \vect{d} $ is the incidence
   direction  $ \vect{d} = (\cos(\theta^{\textrm{inc}}),\sin(\theta^{\textrm{inc}}))^T$ and $\theta^{\textrm{inc}}$ is the scattering angle. 
Because of the symmetry of the problem, we fix the incidence direction to $ \vect{d} = (1,0)^T$
and the scatterer as the unit disk with $ R_0 =1$. Similar to the previous example,
 the second-order Bayliss-Turkel ABC (\ref{BGTABC}) is placed on the circle with radius $R_1 =2$. We consider the following exact solution \cite{Kechroud2009} to analyze
 the pollution and approximation errors and   to avoid  the domain truncation error
 \begin{equation}
u^{\textrm{ex}} = \sum_{m \in \mathbb{Z}}  (i)^m u_{m}^{\textrm{ex}},
\end{equation}
 where $u_{m}^{\textrm{ex}}$ is given by Eq. (\ref{eq:Disk_exact}). To get an accurate reference solution, we truncate
 the above series expansion by summing up on $m$ from $-m^{\textrm{max}}$ to $m^{\textrm{max}}:=[kR_{0}]+30$
 (where $[r]$ denotes the integer part of a real-valued positive number $r$). We report on Fig. \ref{fig:Disk_pw} the real parts of $u^{\textrm{ex}}$ (see Fig. \ref{fig:Disk_imag_pw}) and $u_{h}$ (see Fig. \ref{fig:Disk_real_pw}), where the wavenumber is $k=40$. For IGA, the order of the method is $p=3$ for a density of discretization points per wavelength $n_{\lambda}=10$. 
We clearly see that the wavefield is accurately computed since the absolute error $|u_{m}^{\textrm{ex}}-u_{h}|$ is below $10^{-4}$ as it can be deduced from Fig. \ref{Fig17aa}.

\begin{figure}[t!]
    \centering
     \begin{subfigure}[t]{\textwidth}
        \includegraphics[width=\textwidth]{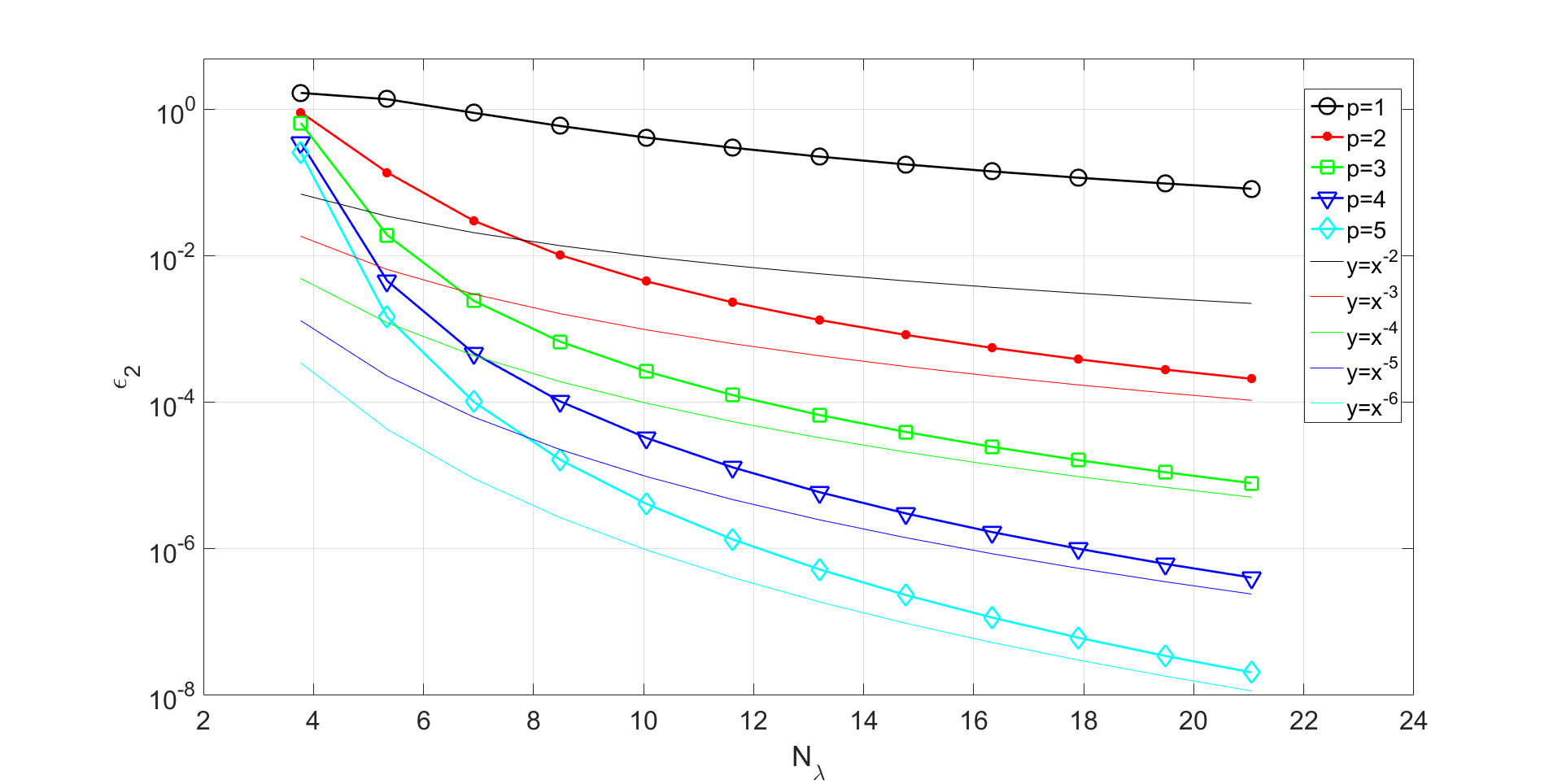}
    \end{subfigure} 
   \caption{Evolution of the error $\epsilon_2$ vs. the discretization density $n_{\lambda}$ for  $k=40$ and $p=1,...,5$.}
\label{fig:Disk_pw_conv}
\end{figure}

\begin{figure}[b!]
    \centering
     \begin{subfigure}[t]{\textwidth}
        \includegraphics[width=\textwidth]{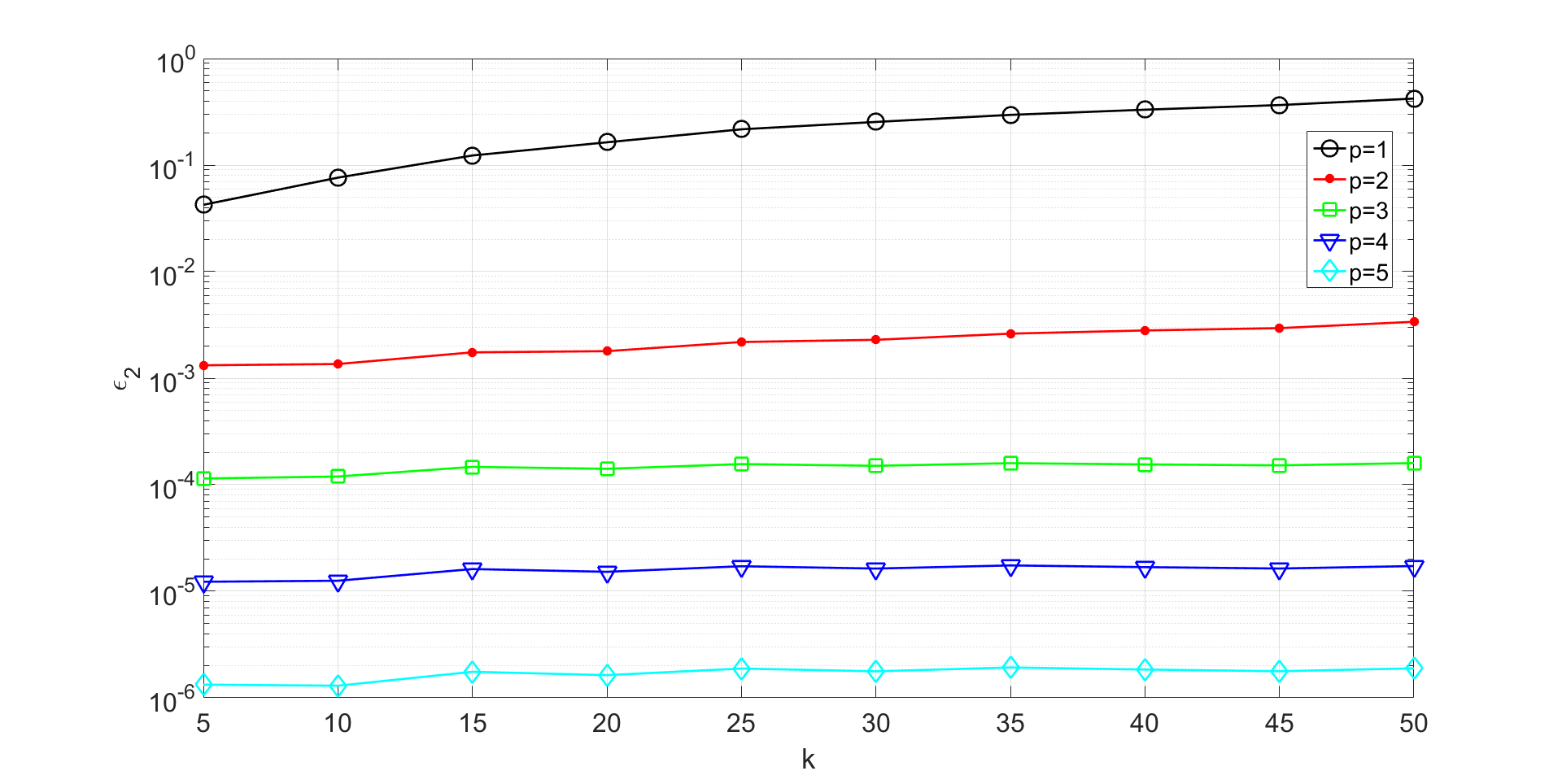}
    \end{subfigure} 
   \caption{Evolution of the error $\epsilon_2$ vs. the wavenumber $k$, for $n_{\lambda}=10$ and $p=1,...,5$.}
\label{fig:Disk_plane_wave_conv_k}
\end{figure}

The evolution of the error with respect to the discretization density $n_{\lambda}$ for $p=1, \cdots, 5$, is shown in Fig. \ref{fig:Disk_pw_conv} where the wavenumber $k$ is equal to $40$.  This shows that if $p$ is larger than $3$, then the $\epsilon_2$-error starts being small (e.g. less than $10^{-2}$) even
for small densities $n_{\lambda}$  (typically  $n_{\lambda}\geq6$). In particular, the error gap between $p=2$ and $p=3$ is important.
In addition, for $n_{\lambda}=5$, we can see on Fig. \ref{Fig17bb} that the error remains
 acceptable but of the order of $2-3\%$, where the maximal
error is located where the amplitude of the wavefield is larger and the solution oscillates.

The evolution of the error $\epsilon_2$ according to the wavenumber $k$ is depicted in Fig. \ref{fig:Disk_plane_wave_conv_k}. For this test,
we fix  $n_{\lambda} = 10$ and  $p=3$ for IGA. As we can see, the error almost does not depend on $k$, which means that the pollution
error is very small. Again, a very good accuracy is obtained for $p\geq3$, the $p$-th order of the method being then visible for $p$
larger than $3$. The dependence with respect to $k$ seems more present for the lowest orders IGA approximations, i.e. $p=1$ and $p=2$.

\begin{figure}[t!]
    \centering
    \begin{subfigure}[t]{0.49\textwidth}
        \includegraphics[width=\textwidth]{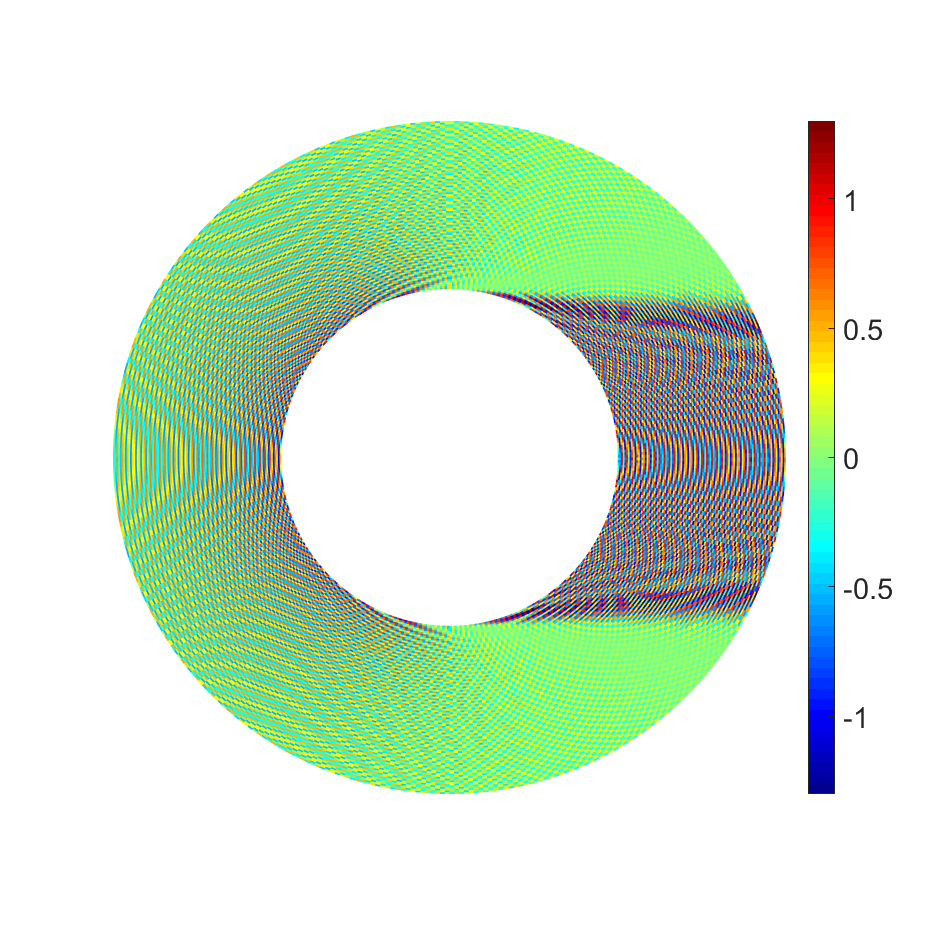}
        \caption{$\Re (u^{\textrm{ex}})$}
        \label{fig:k200n5p3_Real_exact}
    \end{subfigure} 
        \begin{subfigure}[t]{0.49\textwidth}
        \includegraphics[width=\textwidth]{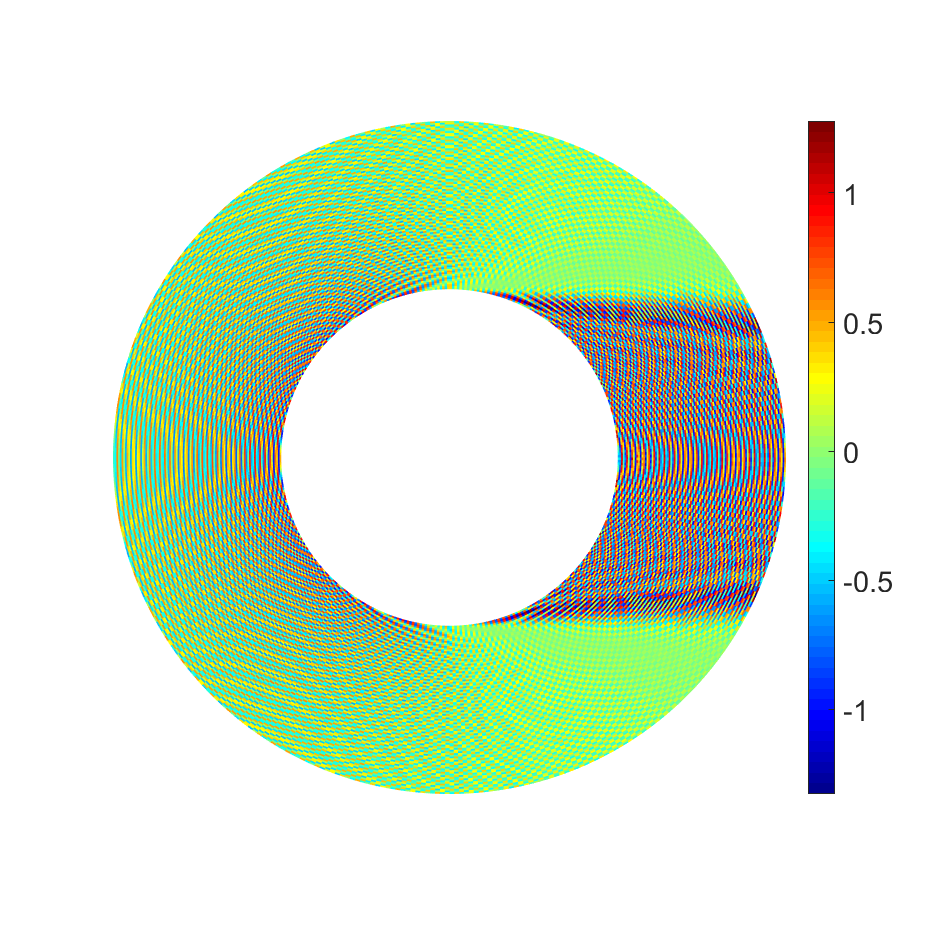}
        \caption{$\Re(u_h)$}
        \label{fig:k200n5p3_Real_num}
    \end{subfigure}
         \begin{subfigure}[t]{0.5\textwidth}
        \includegraphics[width=\textwidth]{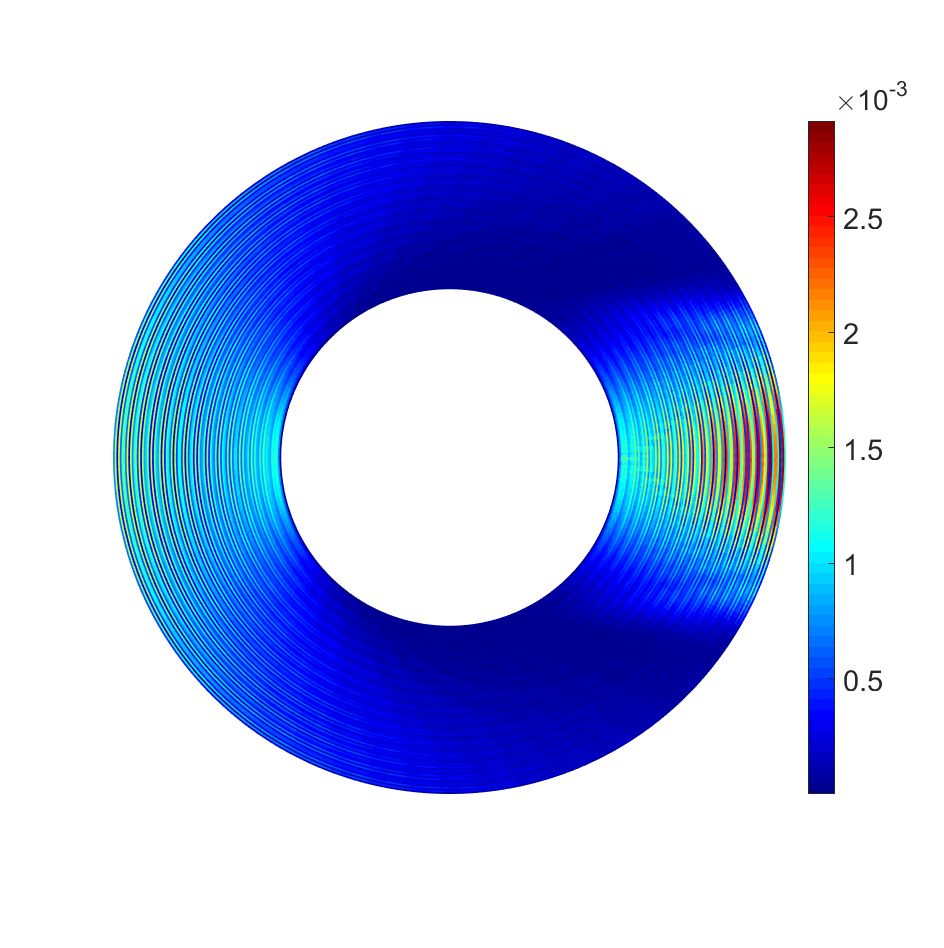}
           \caption{$|u^{\textrm{ex}}-u_h |$}
\label{fig:k200n5p3_errorb}
    \end{subfigure} 
\caption{Comparing the real parts of (a) the exact solution $u^{\textrm{ex}}$ and (b) the numerical IGA solution $u_{h}$,
and (c) absolute error $|u^{\textrm{ex}}-u_h |$,  for $k = 200$, $p=3$, and $n_{\lambda } =5$. }
\label{fig:k200n5p3_real_comp}
\end{figure}

We  consider now a much higher wavenumber, i.e. $k=200$. We report on Fig. \ref{fig:k200n5p3_Real_num} the real part of 
the numerical IGA
solution for $p=3$ and the low density of discretization points per wavelength $n_{\lambda}=5$. The solution
can be compared with the exact solution available on Fig. \ref{fig:k200n5p3_Real_exact}. We immediately
see that the two solutions are the same. This is confirmed on Fig. \ref{fig:k200n5p3_errorb} where
we report the absolute error $|u^{\textrm{ex}}-u_h |$ between the two solutions. This error is of the order of $10^{-3}$,
which is very small.

\begin{figure}[t!]
    \centering
       \begin{subfigure}[t]{0.49\textwidth}
        \includegraphics[width=\textwidth]{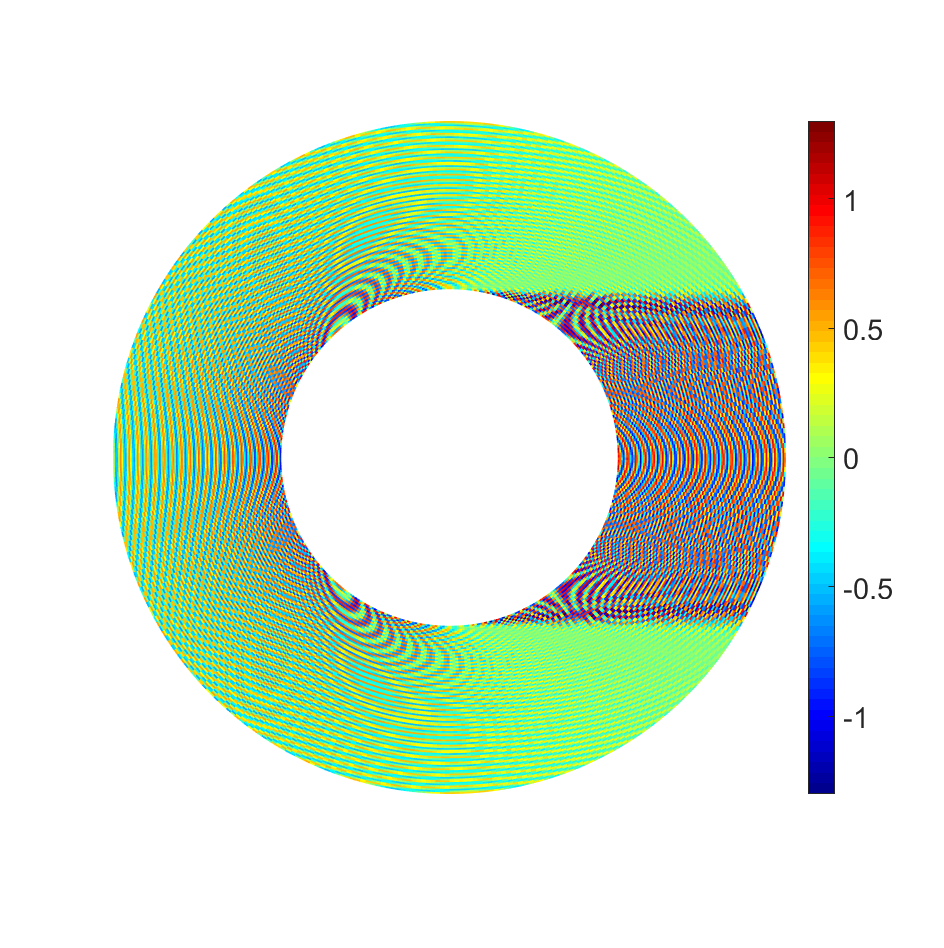}
        \caption{$\Re (u^{\textrm{ex}})$}
        \label{fig:k500n5p3_Real_exact}
    \end{subfigure} 
    \begin{subfigure}[t]{0.49\textwidth}
        \includegraphics[width=\textwidth]{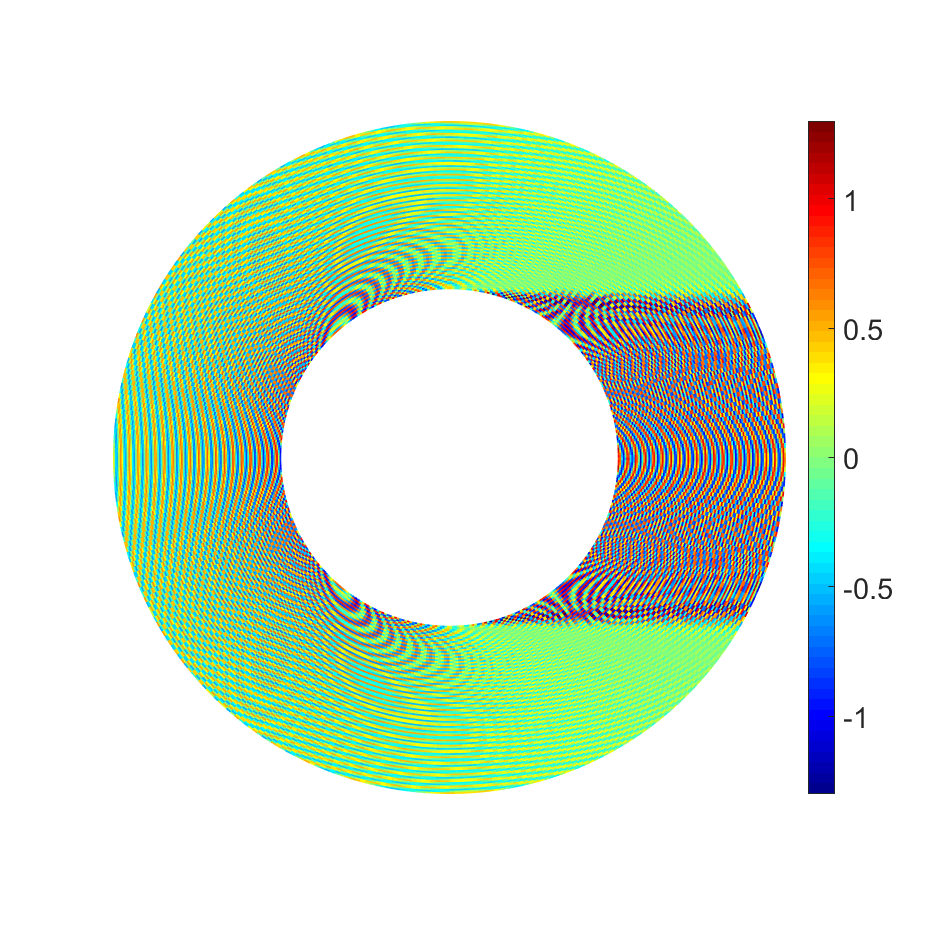}
        \caption{$\Re(u_h)$}
        \label{fig:k500n5p3_Real_num}
    \end{subfigure}
     \begin{subfigure}[t]{0.49\textwidth}
     \includegraphics[width=\textwidth]{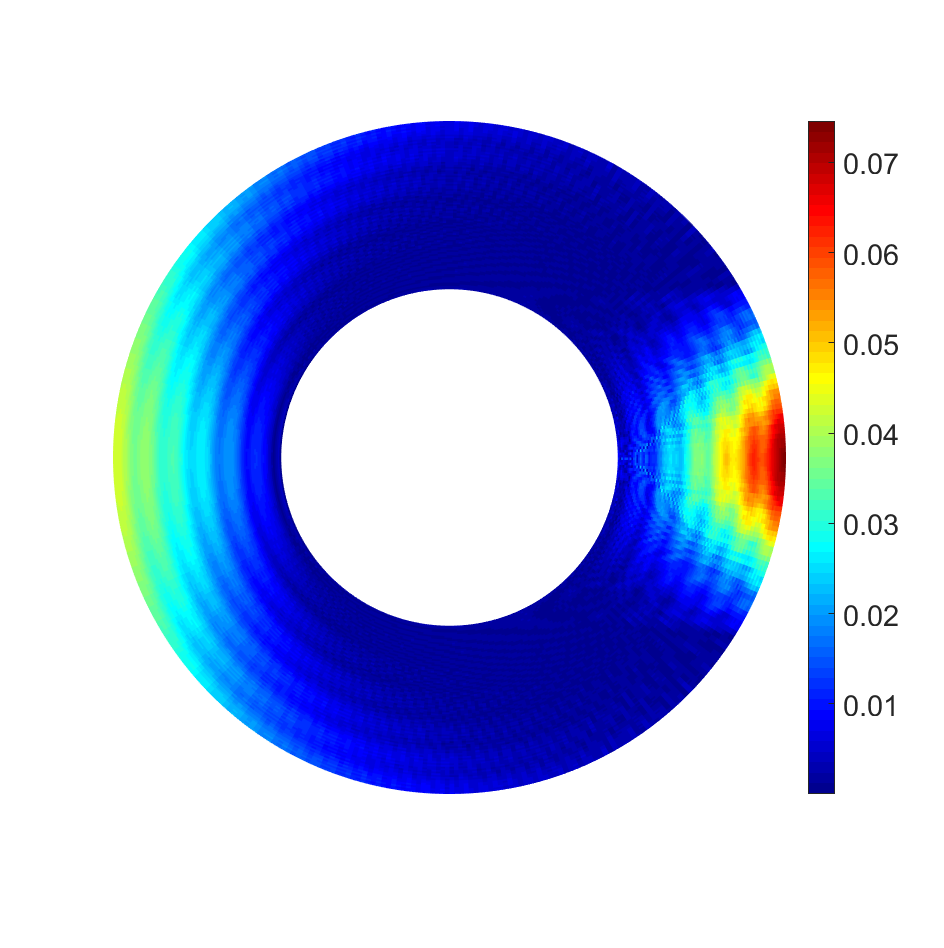}
          \caption{$|u^{\textrm{ex}}-u_h |$ ($p=3$; $n_{\lambda } =5$)}
     \label{fig:k500n5p3_error}
    \end{subfigure} 
         \begin{subfigure}[t]{0.49\textwidth}
        \includegraphics[width=\textwidth]{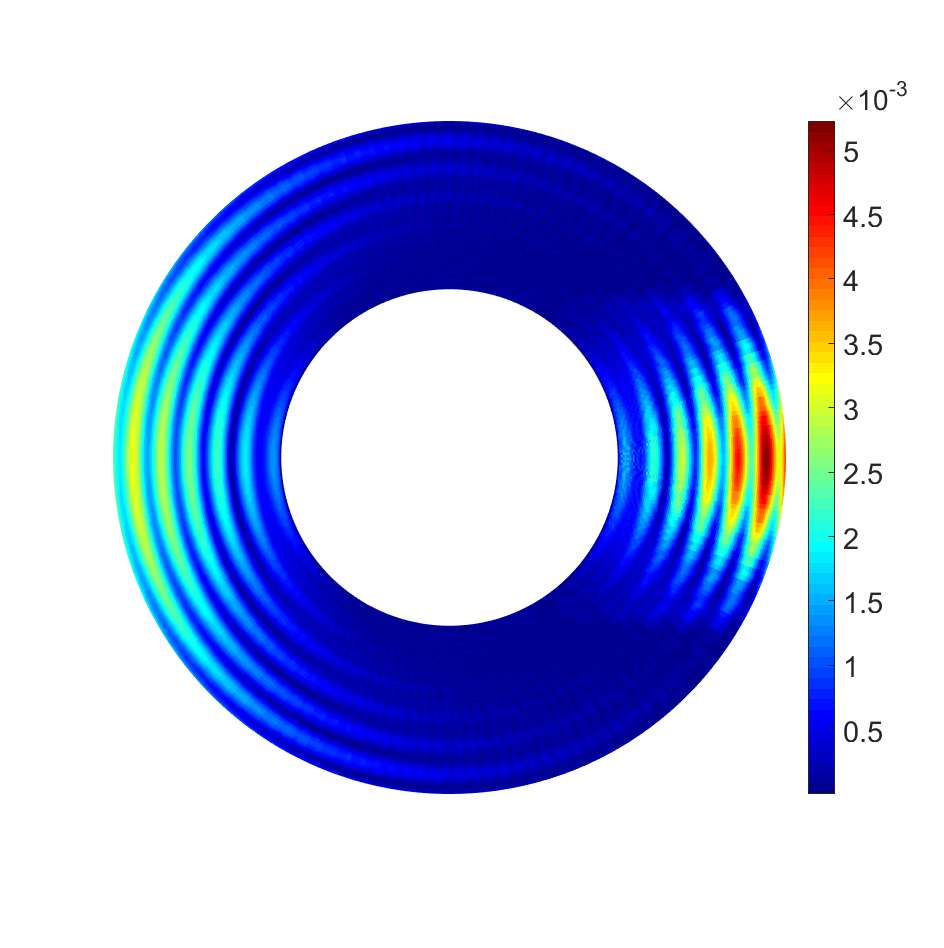}
                \caption{$|u^{\textrm{ex}}-u_h |$ ($p=4$; $n_{\lambda } =5$)}
                             \label{fig:k500n5p4_error}
    \end{subfigure} 
 \caption{Comparing the real parts of (a)  the exact solution $u^{\textrm{ex}}$   
  and (b) the numerical IGA solution $u_{h}$  for $k = 500$, $p=3$, and $n_{\lambda } =5$. Absolute error $|u^{\textrm{ex}}-u_h |$
  for (c) $p=3$; $n_{\lambda } =5$, (d) $p=4$; $n_{\lambda } =5$.
  }
\label{fig:k500n5p3_real_comp}
\end{figure}

Finally, we report now the results for a very high wavenumber $k=500$ on Fig. \ref{fig:k500n5p3_real_comp}. We provide the
real parts of the exact solution $u^{\textrm{ex}}$ (Fig. \ref{fig:k500n5p3_Real_exact}) and
 IGA numerical solution  $u_{h}$ (Fig. \ref{fig:k500n5p3_Real_num}). For IGA, the order of the method is $p=3$ for a density of discretization points per wavelength $n_{\lambda}=5$.
Even if there is no visible difference, plotting the absolute error on Fig. \ref{fig:k500n5p3_error} shows 
 an order of about $6\%$. This can be made much smaller by increasing the order of approximation to $p=4$ 
 by keeping $n_{\lambda}=5$ as seen on Fig. \ref{fig:k500n5p4_error}, leading to an error of about $10^{-3}$.
\section{Conclusions}\label{SectionConclude}
The exterior Helmholtz problem was solved for one-, and two-dimensional problems using IGA. An absorbing boundary condition is applied on a fictitious boundary to truncate the infinite space. This truncation introduces an error in the numerical solution. To better study the performance of IGA, the truncation error was included in the analytical solution so that the pollution error is distinguishable. It is shown that IGA is less prone to pollution error even for large wavenumbers. Also, the pollution error was considerably reduced by increasing the order of the basis functions which can be conveniently done in IGA, even for small densities of discretization points per wavelenght. The smooth representation of the boundaries in IGA and the low pollution error makes IGA a suitable platform for scattering analysis, in particular for prospecting high frequency problems. For high wave numbers it will be interesting to investigate the use of different spaces for geometry and field approximations. Low order geometry description can be used with high order field approximations to investigate super and sub-geometric approaches, also known as GIFT \cite{Xu14}. 







\section{References}
\bibliographystyle{elsarticle-num} 
\small{
\bibliography{Ref_Ac}

\def\cprime{$'$}
\begin{thebibliography}{}

\bibitem[Bou, 1999]{Bouillard1999}
 (1999).
\newblock Error estimation and adaptivity for the finite element method in
  acoustics: 2d and 3d applications.
\newblock {\em Computer Methods in Applied Mechanics and Engineering},
  176(1):147 -- 163.

\bibitem[Antoine et~al., 1999]{AntoineBarucqBendali}
Antoine, X., Barucq, H., and Bendali, A. (1999).
\newblock Bayliss-{T}urkel-like radiation conditions on surfaces of arbitrary
  shape.
\newblock {\em J. Math. Anal. Appl.}, 229(1):184--211.

\bibitem[Antoine and Darbas, 2005]{AntoineDarbasQJMAM}
Antoine, X. and Darbas, M. (2005).
\newblock Alternative integral equations for the iterative solution of acoustic
  scattering problems.
\newblock {\em Quarterly Journal of Mechanics and Applied Mathematics},
  1(58):107--128.

\bibitem[Antoine and Darbas, 2007]{AntoineDarbasESAIM}
Antoine, X. and Darbas, M. (2007).
\newblock Generalized {C}ombined {F}ield {I}ntegral {E}quations for the
  iterative solution of the three-dimensional {H}elmholtz equation.
\newblock {\em Mathematical Modelling and Numerical Analysis}, 1(41):147--167.

\bibitem[Antoine and Geuzaine, 2009]{AntoineCG2009}
Antoine, X. and Geuzaine, C. (2009).
\newblock Phase reduction models for improving the accuracy of the finite
  element solution of time-harmonic scattering problems i: General approach and
  low-order models.
\newblock {\em Journal of Computational Physics}, 228(8):3114 -- 3136.

\bibitem[Astley and Eversman, 1983]{ASTLEY1983}
Astley, R. and Eversman, W. (1983).
\newblock Finite element formulations for acoustical radiation.
\newblock {\em Journal of Sound and Vibration}, 88(1):47 -- 64.

\bibitem[Astley, 1983]{Astley1983-1}
Astley, R.~J. (1983).
\newblock Wave envelope and infinite elements for acoustical radiation.
\newblock {\em International Journal for Numerical Methods in Fluids},
  3(5):507--526.

\bibitem[Babuska and Sauter, 1997]{Babuska1997}
Babuska, I. and Sauter, S. (1997).
\newblock Is the pollution effect of the fem avoidable for the helmholtz
  equation considering high wave numbers?
\newblock {\em SIAM Journal on Numerical Analysis}, 34(6):2392--2423.

\bibitem[Baumeister, 1974]{baumeister1974}
Baumeister, K. (1974).
\newblock {\em Analysis of Sound Propagation in Ducts Using the Wave Envelope
  Concept}.
\newblock NASA technical note. National Aeronautics and Space Administration.

\bibitem[Baumeister et~al., 1977]{baumeister1977}
Baumeister, K., Aeronautics, U. S.~N., Scientific, S.~A., and Office, T.~I.
  (1977).
\newblock {\em Finite-difference theory for sound propagation in a lined duct
  with uniform flow using the wave envelope concept}.
\newblock NASA technical paper. National Aeronautics and Space Administration,
  Scientific and Technical Information Office.

\bibitem[Bayliss and Turkel, 1980]{Turkel3}
Bayliss, A. and Turkel, E. (1980).
\newblock Radiation boundary conditions for wave-like equations.
\newblock {\em Comm. Pure Appl. Math.}, 33(6):707--725.

\bibitem[Bebendorf, 2000]{Bebendorf}
Bebendorf, M. (2000).
\newblock Approximation of boundary element matrices.
\newblock {\em Numerische Mathematik}, 86(4):565--589.

\bibitem[B\'erenger, 1994]{BerengerPML}
B\'erenger, J.-P. (1994).
\newblock A perfectly matched layer for the absorption of electromagnetic
  waves.
\newblock {\em J. Comput. Phys.}, 114(2):185--200.

\bibitem[Bermudez et~al., 2007]{BermudezPML}
Bermudez, A., Hervella-Nieto, L., Prieto, A., and Rodriguez, R. (2007).
\newblock An optimal perfectly matched layer with unbounded absorbing function
  for time-harmonic acoustic scattering problems.
\newblock {\em Journal of Computational Physics}, 223(2):469 -- 488.

\bibitem[Berm{\'u}dez et~al., 2010]{BermudezReview}
Berm{\'u}dez, A., Hervella-Nieto, L., Prieto, A., and Rodr{\'i}guez, R. (2010).
\newblock Perfectly matched layers for time-harmonic second order elliptic
  problems.
\newblock {\em Archives of Computational Methods in Engineering},
  17(1):77--107.

\bibitem[Boubendir et~al., 2012]{BoubendirAntoineGeuzaine}
Boubendir, Y., Antoine, X., and Geuzaine, C. (2012).
\newblock A quasi-optimal non-overlapping domain decomposition algorithm for
  the {H}elmholtz equation.
\newblock {\em Journal of Computational Physics}, 2(231):262--280.

\bibitem[Bouillard and Ihlenburg, 1999]{BouillardIhlenburg99}
Bouillard, P. and Ihlenburg, F. (1999).
\newblock Error estimation and adaptivity for the finite element method in
  acoustics: 2d and 3d applications.
\newblock {\em Computer Methods in Applied Mechanics and Engineering},
  176(1-4):147--163.
\newblock Workshop on New Advances in Adaptive Computational methods, Cachan,
  France, SEP 17-19, 1997.

\bibitem[Bériot et~al., 2016]{Beriot16}
Bériot, H., Prinn, A., and Gabard, G. (2016).
\newblock Efficient implementation of high-order finite elements for helmholtz
  problems.
\newblock {\em International Journal for Numerical Methods in Engineering},
  106(3):213--240.

\bibitem[Bruno et~al., 2004]{Bruno2004}
Bruno, O.~P., Geuzaine, C.~A., Monro, J.~A., and Reitich, F. (2004).
\newblock Prescribed error tolerances within fixed computational times for
  scattering problems of arbitrarily high frequency: the convex case.
\newblock {\em Philosophical Transactions of the Royal Society of London A:
  Mathematical, Physical and Engineering Sciences}, 362(1816):629--645.

\bibitem[Cessenat and Despres, 1998]{CessenatDespresUltraWeak}
Cessenat, O. and Despres, B. (1998).
\newblock Application of an ultra weak variational formulation of elliptic pdes
  to the two-dimensional helmholtz problem.
\newblock {\em SIAM Journal on Numerical Analysis}, 35(1):255--299.

\bibitem[Chandler-Wilde et~al., 2012]{ChandlerWildeReview}
Chandler-Wilde, S.~N., Graham, I.~G., Langdon, S., and Spence, E.~A. (2012).
\newblock Numerical-asymptotic boundary integral methods in high-frequency
  acoustic scattering.
\newblock {\em Acta Numer.}, 21:89--305.

\bibitem[Chew and Weedon, 1994]{ChewPML}
Chew, W. and Weedon, W. ({1994}).
\newblock {A 3D Perfectly matched medium from modified {M}axwell's equations
  with stretched coordinates}.
\newblock {\em {Microwave and Optical Technology Letters}},
  {7}({13}):{599--604}.

\bibitem[Colton and Kress, 1993]{colton1983integral}
Colton, D. and Kress, R. (1993).
\newblock {\em Integral equation methods in scattering theory}.
\newblock Pure and applied mathematics. Wiley.

\bibitem[Coox et~al., 2016a]{Coox2016}
Coox, L., Atak, O., Vandepitte, D., and Desmet, W. (2016a).
\newblock An isogeometric indirect boundary element method for solving acoustic
  problems in open-boundary domains.
\newblock {\em Computer Methods in Applied Mechanics and Engineering},
  pages~--.

\bibitem[Coox et~al., 2016b]{Coox2016441}
Coox, L., Deckers, E., Vandepitte, D., and Desmet, W. (2016b).
\newblock A performance study of nurbs-based isogeometric analysis for interior
  two-dimensional time-harmonic acoustics.
\newblock {\em Computer Methods in Applied Mechanics and Engineering}, 305:441
  -- 467.

\bibitem[Cottrell et~al., 2006]{Cottrell20065257}
Cottrell, J., Reali, A., Bazilevs, Y., and Hughes, T. (2006).
\newblock Isogeometric analysis of structural vibrations.
\newblock {\em Computer Methods in Applied Mechanics and Engineering},
  195(41–43):5257 -- 5296.
\newblock John H. Argyris Memorial Issue. Part \{II\}.

\bibitem[Cottrell et~al., 2009]{Cottrell2009}
Cottrell, J.~A., Hughes, T. J.~R., and Bazilevs, Y. (2009).
\newblock {\em Isogeometric Analysis: Toward Integration of CAD and FEA}.
\newblock Wiley Publishing, 1st edition.

\bibitem[Darbas et~al., 2013]{DDL}
Darbas, M., Darrigrand, E., and Lafranche, Y. (2013).
\newblock Combining analytic preconditioner and fast multipole method for the
  3-{D} {H}elmholtz equation.
\newblock {\em J. Comput. Phys.}, 236:289--316.

\bibitem[Darve, 2000]{Darve}
Darve, E. (2000).
\newblock The fast multipole method: numerical implementation.
\newblock {\em J. Comput. Phys.}, 160(1):195--240.

\bibitem[de~Falco et~al., 2011]{deFalco2011}
de~Falco, C., Reali, A., and Vázquez, R. (2011).
\newblock Geopdes: A research tool for isogeometric analysis of \{PDEs\}.
\newblock {\em Advances in Engineering Software}, 42(12):1020 -- 1034.

\bibitem[Despr{\'e}s, 1990]{despres:90}
Despr{\'e}s, B. ({1990}).
\newblock {D\'ecomposition de domaine et probl\`eme de {H}elmholtz}.
\newblock {\em {C.R. Acad. Sci. Paris}}, {1}({6}):{313--316}.

\bibitem[Despr{\'e}s et~al., 1992]{despres-etal:92}
Despr{\'e}s, B., Joly, P., and Roberts, J.~E. ({1992}).
\newblock {A domain decomposition method for the harmonic {M}axwell equations}.
\newblock In {\em {Iterative methods in linear algebra (Brussels, 1991)}},
  pages {475--484}, {Amsterdam}. {North-Holland}.

\bibitem[Farhat et~al., 2001]{FarhatEnriched}
Farhat, C., Harari, I., and Franca, L. ({2001}).
\newblock {The discontinuous enrichment method}.
\newblock {\em Computer Methods in Applied Mechanics and Engineering},
  {190}({48}):{6455--6479}.

\bibitem[Fusseder et~al., 2015]{Fußeder2015}
Fusseder, D., Simeon, B., and Vuong, A.-V. (2015).
\newblock Fundamental aspects of shape optimization in the context of
  isogeometric analysis.
\newblock {\em Computer Methods in Applied Mechanics and Engineering}, 286:313
  -- 331.

\bibitem[Gabard et~al., 2011a]{GabardGamallo}
Gabard, G., Gamallo, P., and Huttunen, T. ({2011}a).
\newblock {A comparison of wave-based discontinuous Galerkin, ultra-weak and
  least-square methods for wave problems}.
\newblock {\em International Journal for Numerical Methods in Engineering},
  {85}({3}):{380--402}.

\bibitem[Gabard et~al., 2011b]{GabardGamallo11}
Gabard, G., Gamallo, P., and Huttunen, T. (2011b).
\newblock International journal for numerical methods in engineering.
\newblock {\em International Journal for Numerical Methods in Engineering},
  85(3):380--402.

\bibitem[Gander et~al., 2002]{GanderMagouNataf}
Gander, M., Magoules, F., and Nataf, F. ({2002}).
\newblock {Optimized Schwarz methods without overlap for the Helmholtz
  equation}.
\newblock {\em SIAM J. Scientific Computing}, {24}({1}):{38--60}.

\bibitem[Geuzaine et~al., 2008]{Geuzaine2008}
Geuzaine, C., Bedrossian, J., and Antoine, X. (2008).
\newblock An amplitude formulation to reduce the pollution error in the finite
  element solution of time-harmonic scattering problems.
\newblock {\em IEEE Transactions on Magnetics}, 44(6):782--785.

\bibitem[Giladi and Keller, 2001]{GILADI2001}
Giladi, E. and Keller, J. (2001).
\newblock A hybrid numerical asymptotic method for scattering problems.
\newblock {\em Journal of Computational Physics}, 174(1):226 -- 247.

\bibitem[Giorgiani et~al., 2013]{GiorgianiModesto13}
Giorgiani, G., Modesto, D., Fernandez-Mendez, S., and Huerta, A. ({2013}).
\newblock High-order continuous and discontinuous galerkin methods for wave
  problems.
\newblock {\em {International Journal for Numerical Methods in Fluids}},
  {73}({10}):{883--903}.

\bibitem[Givoli, 2004]{ABCReview3}
Givoli, D. ({2004}).
\newblock {High-order local non-reflecting boundary conditions: a review}.
\newblock {\em {Wave Motion}}, {39}({4}):{319--326}.

\bibitem[Hagstrom, 2003]{ABCReview2}
Hagstrom, T. ({2003}).
\newblock {New results on absorbing layers and radiation boundary conditions}.
\newblock In {Ainsworth, M and Davies, P and Duncan, D and Martin, P and Rynne,
  B}, editor, {\em {Topics in Computational Wave Propagation: Direct and
  Inverse Problems}}, volume~{31} of {\em {Lecture Notes in Computational
  Science and Engineering}}, pages {1--42}.
\newblock {Symposium on Computational Methods for Wave Propagation in Direct
  Scattering, Univ Durham, Durham, England, Jul 15-25, 2002}.

\bibitem[Harari and Magoules, 2004]{HarariMagoules}
Harari, I. and Magoules, F. ({2004}).
\newblock {Numerical investigations of stabilized finite element computations
  for acoustics}.
\newblock {\em {Wave Motion}}, {39}({4}):{339--349}.

\bibitem[Hughes et~al., 2005]{Hughes2005}
Hughes, T., Cottrell, J., and Bazilevs, Y. (2005).
\newblock Isogeometric analysis: Cad, finite elements, nurbs, exact geometry
  and mesh refinement.
\newblock {\em Computer Methods in Applied Mechanics and Engineering},
  194(39–41):4135 -- 4195.

\bibitem[Hughes et~al., 2008]{Hughes20084104}
Hughes, T., Reali, A., and Sangalli, G. (2008).
\newblock Duality and unified analysis of discrete approximations in structural
  dynamics and wave propagation: Comparison of p-method finite elements with
  k-method \{NURBS\}.
\newblock {\em Computer Methods in Applied Mechanics and Engineering},
  197(49–50):4104 -- 4124.

\bibitem[Hughes et~al., 2014]{Hughes2014290}
Hughes, T.~J., Evans, J.~A., and Reali, A. (2014).
\newblock Finite element and \{NURBS\} approximations of eigenvalue,
  boundary-value, and initial-value problems.
\newblock {\em Computer Methods in Applied Mechanics and Engineering}, 272:290
  -- 320.

\bibitem[Huttunen et~al., 2009]{Huttunen2008}
Huttunen, T., Gamallo, P., and Astley, R.~J. (2009).
\newblock Comparison of two wave element methods for the helmholtz problem.
\newblock {\em Communications in Numerical Methods in Engineering},
  25(1):35--52.

\bibitem[Ihlenburg, 1998]{Ihlenburg1998}
Ihlenburg, F. (1998).
\newblock {\em Finite Element Analysis of Acoustic Scattering}.
\newblock Springer.

\bibitem[Ihlenburg and Babuška, 1995]{Ihlenburg1995}
Ihlenburg, F. and Babuška, I. (1995).
\newblock Finite element solution of the helmholtz equation with high wave
  number part i: The h-version of the fem.
\newblock {\em Computers and Mathematics with Applications}, 30(9):9--37.

\bibitem[Ihlenburg and Babuska, 1997a]{IhlenburgBabuskaI97}
Ihlenburg, F. and Babuska, I. ({1997}a).
\newblock {Finite element solution of the Helmholtz equation with high wave
  number .2. The h-p version of the FEM}.
\newblock {\em {SIAM Journal on Numerical Analysis}}, {34}({1}):{315--358}.

\bibitem[Ihlenburg and Babuska, 1997b]{Ihlenburg1997}
Ihlenburg, F. and Babuska, I. (1997b).
\newblock Finite element solution of the helmholtz equation with high wave
  number part ii: The h-p version of the fem.
\newblock {\em SIAM J. Numer. Anal.}, 34(1):315--358.

\bibitem[Kechroud et~al., 2005]{AntoineABC}
Kechroud, R., Antoine, X., and Soula{\"{\i}}mani, A. (2005).
\newblock Numerical accuracy of a {P}ad\'e-type non-reflecting boundary
  condition for the finite element solution of acoustic scattering problems at
  high-frequency.
\newblock {\em Internat. J. Numer. Methods Engrg.}, 64(10):1275--1302.

\bibitem[Kechroud et~al., 2009]{Kechroud2009}
Kechroud, R., Soulaimani, A., and Antoine, X. (2009).
\newblock A performance study of plane wave finite element methods with a
  padé-type artificial boundary condition in acoustic scattering.
\newblock {\em Advances in Engineering Software}, 40(8):738 -- 750.

\bibitem[Kostas et~al., 2015]{Kostas2015}
Kostas, K., Ginnis, A., Politis, C., and Kaklis, P. (2015).
\newblock Ship-hull shape optimization with a t-spline based bem–isogeometric
  solver.
\newblock {\em Computer Methods in Applied Mechanics and Engineering}, 284:611
  -- 622.

\bibitem[Laghrouche and Bettess, 2000]{LAGHROUCHE2000}
Laghrouche, O. and Bettess, P. (2000).
\newblock Short wave modelling using special finite elements.
\newblock {\em Journal of Computational Acoustics}, 08(01):189--210.

\bibitem[Lian et~al., 2016]{Lian2016}
Lian, H., Kerfriden, P., and Bordas, S. P.~A. (2016).
\newblock Implementation of regularized isogeometric boundary element methods
  for gradient-based shape optimization in two-dimensional linear elasticity.
\newblock {\em International Journal for Numerical Methods in Engineering},
  106(12):972--1017.

\bibitem[Lian et~al., 2012]{Lian2013}
Lian, H., Simpson, R., and Bordas, S. (2012).
\newblock Stress analysis without meshing: Isogeometric boundary-element
  method.
\newblock {\em Proceedings of the Institution of Civil Engineers: Engineering
  and Computational Mechanics}, 166:88 -- 89.

\bibitem[Lian et~al., 2013]{Lian2013R}
Lian, H., Simpson, R., and Bordas, S. P.~A. (2013).
\newblock Sensitivity analysis and shape optimization through a t-spline
  isogeometric boundary element method.
\newblock {\em International Conference on Computational Mechanics}.

\bibitem[Liu, 2009]{FMMBook}
Liu, Y. (2009).
\newblock {\em Fast Multipole Boundary Element Method. Theory and Applications
  in Engineering}.
\newblock Cambridge University Press.

\bibitem[Medvinsky and Turkel, 2010]{Turkel1}
Medvinsky, M. and Turkel, E. (2010).
\newblock On surface radiation conditions for an ellipse.
\newblock {\em J. Comput. Appl. Math.}, 234(6):1647--1655.

\bibitem[N{\'e}d{\'e}lec, 2001]{NedelecBook}
N{\'e}d{\'e}lec, J.-C. (2001).
\newblock {\em Acoustic and electromagnetic equations}, volume 144 of {\em
  Applied Mathematical Sciences}.
\newblock Springer-Verlag, New York.
\newblock Integral representations for harmonic problems.

\bibitem[Nguyen et~al., 2015]{Nguyen2015}
Nguyen, V.~P., Anitescu, C., Bordas, S.~P., and Rabczuk, T. (2015).
\newblock Isogeometric analysis: An overview and computer implementation
  aspects.
\newblock {\em Mathematics and Computers in Simulation}, 117:89 -- 116.

\bibitem[Ortiz and Sanchez, 2001]{Ortiz2001}
Ortiz, P. and Sanchez, E. (2001).
\newblock An improved partition of unity finite element model for diffraction
  problems.
\newblock {\em International Journal for Numerical Methods in Engineering},
  50(12):2727--2740.

\bibitem[Piegl and Tiller, 1997]{Piegl97}
Piegl, L. and Tiller, W. (1997).
\newblock {\em Introduction to finite element analysis : formulation,
  verification and validation}.
\newblock Springer.

\bibitem[Saad, 2003]{SaadBook}
Saad, Y. (2003).
\newblock {\em Iterative methods for sparse linear systems}.
\newblock Society for Industrial and Applied Mathematics, Philadelphia, PA,
  second edition.

\bibitem[Scott et~al., 2013]{Scott2013}
Scott, M., Simpson, R., Evans, J., Lipton, S., Bordas, S., Hughes, T., and
  Sederberg, T. (2013).
\newblock Isogeometric boundary element analysis using unstructured t-splines.
\newblock {\em Computer Methods in Applied Mechanics and Engineering}, 254:197
  -- 221.

\bibitem[Simpson et~al., 2013]{Simpson2013}
Simpson, R., Bordas, S., Lian, H., and Trevelyan, J. (2013).
\newblock An isogeometric boundary element method for elastostatic analysis: 2d
  implementation aspects.
\newblock {\em Computers and Structures}, 118:2 -- 12.
\newblock Special Issue: \{UK\} Association for Computational Mechanics in
  Engineering.

\bibitem[Simpson et~al., 2012]{Simpson2012}
Simpson, R., Bordas, S., Trevelyan, J., and Rabczuk, T. (2012).
\newblock A two-dimensional isogeometric boundary element method for
  elastostatic analysis.
\newblock {\em Computer Methods in Applied Mechanics and Engineering},
  209–212:87 -- 100.

\bibitem[{Thompson}, 2006]{Thompson2006}
{Thompson}, L.~L. (2006).
\newblock {A review of finite-element methods for time-harmonic acoustics}.
\newblock {\em Acoustical Society of America Journal}, 119:1315.

\bibitem[Thompson and Pinsky, ]{ThompsonPinsky95}
Thompson, L.~L. and Pinsky, P.~M.
\newblock A galerkin least-squares finite element method for the
  two-dimensional helmholtz equation.
\newblock {\em International Journal for Numerical Methods in Engineering},
  38(3).

\bibitem[Tsynkov, 1998]{ABCReview1}
Tsynkov, S. ({1998}).
\newblock {Numerical solution of problems on unbounded domains. A review}.
\newblock {\em {Applied Numerical Mathematics}}, {27}({4}):{465--532}.

\bibitem[Turkel, 2008]{Turkel2}
Turkel, E. (2008).
\newblock Boundary conditions and iterative schemes for the {H}elmholtz
  equation in unbounded region.
\newblock In {\em Computational Methods for Acoustics Problems}, pages
  127--158. Saxe-Coburg Publications.

\bibitem[Turkel et~al., 2004]{Turkel2004}
Turkel, E., Farhat, C., and Hetmaniuk, U. (2004).
\newblock Improved accuracy for the helmholtz equation in unbounded domains.
\newblock {\em International Journal for Numerical Methods in Engineering},
  59(15):1963--1988.

\bibitem[Wall et~al., 2008]{Wolfgang2008}
Wall, W.~A., Frenzel, M.~A., and Cyron, C. (2008).
\newblock Isogeometric structural shape optimization.
\newblock {\em Computer Methods in Applied Mechanics and Engineering},
  197(33–40):2976 -- 2988.

\bibitem[Xu et~al., 2014]{Xu14}
Xu, G., ~, Elena, A., and Bordas, S. P.~A. (2014).
\newblock Geometry-independent field approximation for spline-based finite
  element methods.
\newblock {\em Proceedings of the 11th World Congress in Computational
  Mechanics}.

\bibitem[Zhao et~al., 2005]{ZhaoVouvakis}
Zhao, K., Vouvakis, M.~N., and Lee, J.-F. (2005).
\newblock The adaptive cross approximation algorithm for accelerated method of
  moments computations of emc problems.
\newblock {\em IEEE Transactions on Electromagnetic Compatibility},
  47(4):763--773.

\end{thebibliography}
\biboptions{sort&compress}
}
\end{document}